\documentclass[12pt,epsf]{article}

\usepackage{times}
\usepackage{graphicx}
\usepackage{amsfonts}
\usepackage{amsmath}
\usepackage{amssymb}
\usepackage{amstext}
\usepackage{amsthm}
\usepackage{youngtab}
\usepackage{epsfig}
\usepackage{color}
\psfigdriver{dvips}
\usepackage{latexsym}
\usepackage[matrix,curve]{xypic}
\usepackage{rotating}
\rotdriver{dvips}

\begin{document}

\baselineskip 6mm
\renewcommand{\thefootnote}{\fnsymbol{footnote}}


\newcommand{\nc}{\newcommand}
\newcommand{\rnc}{\renewcommand}


\rnc{\baselinestretch}{1.24}    
\setlength{\jot}{6pt}       
\rnc{\arraystretch}{1.24}   

\makeatletter
\rnc{\theequation}{\thesection.\arabic{equation}}
\@addtoreset{equation}{section}
\makeatother



\nc{\be}{\begin{equation}}

\nc{\ee}{\end{equation}}

\nc{\bea}{\begin{eqnarray}}

\nc{\eea}{\end{eqnarray}}

\nc{\xx}{\nonumber\\}

\nc{\ct}{\cite}

\nc{\la}{\label}

\nc{\eq}[1]{(\ref{#1})}

\nc{\newcaption}[1]{\centerline{\parbox{6in}{\caption{#1}}}}

\nc{\fig}[3]{

\begin{figure}
\centerline{\epsfxsize=#1\epsfbox{#2.eps}}
\newcaption{#3. \label{#2}}
\end{figure}
}


\def\CA{{\cal A}}
\def\CC{{\cal C}}
\def\CD{{\cal D}}
\def\CE{{\cal E}}
\def\CF{{\cal F}}
\def\CG{{\cal G}}
\def\CH{{\cal H}}
\def\CK{{\cal K}}
\def\CL{{\cal L}}
\def\CM{{\cal M}}
\def\CN{{\cal N}}
\def\CO{{\cal O}}
\def\CP{{\cal P}}
\def\CR{{\cal R}}
\def\CS{{\cal S}}
\def\CU{{\cal U}}
\def\CV{{\cal V}}
\def\CW{{\cal W}}
\def\CY{{\cal Y}}
\def\CZ{{\cal Z}}


\def\IB{{\hbox{{\rm I}\kern-.2em\hbox{\rm B}}}}
\def\IC{\,\,{\hbox{{\rm I}\kern-.50em\hbox{\bf C}}}}
\def\ID{{\hbox{{\rm I}\kern-.2em\hbox{\rm D}}}}
\def\IF{{\hbox{{\rm I}\kern-.2em\hbox{\rm F}}}}
\def\IH{{\hbox{{\rm I}\kern-.2em\hbox{\rm H}}}}
\def\IN{{\hbox{{\rm I}\kern-.2em\hbox{\rm N}}}}
\def\IP{{\hbox{{\rm I}\kern-.2em\hbox{\rm P}}}}
\def\IR{{\hbox{{\rm I}\kern-.2em\hbox{\rm R}}}}
\def\IZ{{\hbox{{\rm Z}\kern-.39em\hbox{\rm Z}}}}


\def\a{\alpha}
\def\b{\beta}
\def\d{\delta}
\def\ep{\epsilon}
\def\ga{\gamma}
\def\k{\kappa}
\def\l{\lambda}
\def\s{\sigma}
\def\t{\theta}
\def\w{\omega}
\def\G{\Gamma}


\def\half{\frac{1}{2}}
\def\dint#1#2{\int\limits_{#1}^{#2}}
\def\goto{\rightarrow}
\def\para{\parallel}
\def\brac#1{\langle #1 \rangle}
\def\curl{\nabla\times}
\def\div{\nabla\cdot}
\def\p{\partial}


\def\Tr{{\rm Tr}\,}
\def\det{{\rm det}}


\def\vare{\varepsilon}
\def\zbar{\bar{z}}
\def\wbar{\bar{w}}
\def\what#1{\widehat{#1}}


\def\ad{\dot{a}}
\def\bd{\dot{b}}
\def\cd{\dot{c}}
\def\dd{\dot{d}}
\def\so{SO(4)}
\def\bfr{{\bf R}}
\def\bfc{{\bf C}}
\def\bfz{{\bf Z}}

\begin{titlepage}


\hfill\parbox{3.7cm} {{\tt arXiv:1107.2095}}

\vspace{15mm}

\begin{center}
{\Large \bf  Calabi-Yau Manifolds, Hermitian Yang-Mills Instantons and Mirror Symmetry}

\vspace{10mm}

Hyun Seok Yang ${}^a$ \footnote{hsyang@sogang.ac.kr}
and Sangheon Yun ${}^b$ \footnote{sanhan@ewha.ac.kr}
\\[10mm]

${}^a$ {\sl Center for Quantum Spacetime, Sogang University, Seoul 121-741, Korea}

${}^b$ {\sl Institute for the Early Universe, Ewha Womans University, Seoul 120-750, Korea}

\end{center}

\thispagestyle{empty}

\vskip1cm


\centerline{\bf ABSTRACT}
\vskip 4mm
\noindent

We address the issue why Calabi-Yau manifolds exist with a mirror pair. We observe that
the irreducible spinor representation of the Lorentz group $Spin(6)$ requires us to consider
the vector spaces of two-forms and four-forms on an equal footing.
The doubling of the two-form vector space due to the Hodge duality doubles
the variety of six-dimensional spin manifolds.
We explore how the doubling is related to the mirror symmetry of Calabi-Yau manifolds.
Via the gauge theory formulation of six-dimensional Riemannian manifolds,
we show that the curvature tensor of a Calabi-Yau manifold satisfies the Hermitian Yang-Mills
equations on the Calabi-Yau manifold.
Therefore the mirror symmetry of Calabi-Yau manifolds can be recast as
the mirror pair of Hermitian Yang-Mills instantons.
We discuss the mirror symmetry from the gauge theory perspective. \\




\vspace{1cm}

\today

\end{titlepage}

\renewcommand{\thefootnote}{\arabic{footnote}}
\setcounter{footnote}{0}

\section{Introduction}

String theory predicts \cite{string-book} that six-dimensional Riemannian manifolds
have to play an important role in explaining our four-dimensional world.
They serve as an internal geometry of string theory with six extra dimensions and
their shapes and topology determine a detailed structure of the multiplets
for elementary particles and gauge fields through the Kaluza-Klein compactification.
This program, initiated by a classic paper \cite{vacuum-string}, tries to make contact with
a low-energy phenomenology in our four-dimensional world. In particular, a Calabi-Yau (CY) manifold
ia a (compact) K\"ahler manifold with vanishing Ricci curvature and so a vacuum solution of
the Einstein equations. They have a prominent role in superstring theory and have been a central focus
in both contemporary mathematics and mathematical physics.
As the holonomy group of CY manifolds is $SU(3)$, the compactification onto a CY manifold
in heterotic superstring theory preserves $\mathcal{N}=1$ supersymmetry in four dimensions.
One of the most interesting features in the CY compactification is that
type II superstring theories compactified on two distinct CY manifolds lead to an identical
effective field theory in four dimensions \cite{string-book,mirror-book}.
This suggests that CY manifolds exist with a mirror pair $(M, \widetilde{M})$
where the number of vector multiplets $h^{1,1}(M)$ on $M$ is the same as
the number of hypermultiplets $h^{2,1}(\widetilde{M})$ on $\widetilde{M}$ and vice versa.
Here $h^{p,q}(M) = \mathrm{dim} H^{p,q} (M)$ is the Hodge number of a CY manifold $M$.
This duality between two CY manifolds is known as the mirror symmetry \ct{mirror-book}.
While many beautiful properties of the mirror symmetry have been discovered and it has been even proven
for some cases, it is fair to say that we are still far away from a deep understanding
for the origin of mirror symmetry.

Mirror symmetry is a correspondence between two topologically distinct CY manifolds
that give rise to the exactly same physical theory. To recapitulate the mirror symmetry,
let $M$ be a compact CY manifold. The only non-trivial cohomology
of the CY manifold is contained in $H^{1,1}(M)$ and $H^{2,1}(M)$ besides
the one-dimensional cohomologies $h^{0,0}(M) = h^{3,3}(M) = h^{3,0}(M) = h^{0,3}(M) = 1$.
These cohomology classes parameterize CY moduli.
It is known \ct{mirror-book} that every $H^{1,1}(M)$, on one hand, is represented by a real closed
$(1,1)$-form which forms a K\"ahler class represented by the K\"ahler form
of a CY manifold $M$. The elements in $H^{1,1}(M)$ infinitesimally
change the K\"ahler structure of the CY manifold and are therefore
called K\"ahler moduli. (In string theory, these moduli are usually complexified by including $B$-field.)
On the other hand, $H^{2,1}(M)$ parameterizes the complex structure moduli
of a CY manifold $M$. It is thanks to the fact that the cohomology class
of (2,1)-forms is isomorphic to the cohomology class $H^1_{\overline{\partial}}(TM)$,
the first Dolbeault cohomology group of $M$ with values in a holomorphic tangent
bundle $TM$, that characterizes infinitesimal complex structure deformations.
Hence the mirror symmetry of CY manifolds is the duality between two different
CY 3-folds $M$ and $\widetilde{M}$ such that the Hodge numbers of $M$ and
$\widetilde{M}$ satisfy the relations \ct{mirror-book}
\be \la{mirror-cyh1}
h^{1,1}(M) = h^{2,1}(\widetilde{M}), \qquad  h^{2,1}(M) = h^{1,1}(\widetilde{M}),
\ee
or in a more general form
\be \la{mirror-cyh2}
h^{p,q}(M) = h^{3-p,q}(\widetilde{M}),
\ee
where the Hodge number $h^{p,q}$ of a CY manifold
satisfies the relations $h^{p,q} = h^{q,p}$ and $h^{p,q} = h^{3-p,3-q}$.
As we have mentioned above, the only nontrivial deformations of a CY manifold
are generated by the cohomology classes in $H^{1,1}(M)$ and $H^{2,1}(M)$ where $h^{1,1}(M)$ is the number
of possible (in general, complexified) K\"ahler forms and $h^{2,1}(M)$ is the dimension of the complex
structure moduli space of $M$. Mirror symmetry suggests that for each CY 3-fold $M$ there exists
another CY 3-fold $\widetilde{M}$ whose Hodge numbers obey the relation \eq{mirror-cyh1}.

From a physical point of view, two CY manifolds are related by mirror symmetry if
the corresponding $\mathcal{N} = 2$ superconformal field theories are mirror \cite{witten-mirror}.
Two $\mathcal{N} = 2$ superconformal field theories are said to be mirror if they
are equivalent as quantum field theories. The mirror symmetry was interpreted mathematically
by M. Kontsevich in his 1994 ICM talk as an equivalence of derived categories,
dubbed the homological mirror symmetry \cite{hms}. The homological mirror symmetry states that
the derived category of coherent sheaves on a K\"ahler manifold should be isomorphic
to the Fukaya category of a mirror symplectic manifold \cite{kias}.
The Fukaya category is described by the Lagrangian submanifold of
a given symplectic manifold as its objects and the Floer homology groups as their morphisms.
Hence the homological mirror symmetry formulates the mirror symmetry as an equivalence
between certain aspects of complex geometry of a CY manifold and certain aspects of symplectic geometry
of a mirror CY manifold in all dimensions.
The geometric approach to mirror symmetry was
also unveiled in \cite{syz-mirror} that mirror symmetry is a geometric version of the
Fourier-Mukai transformation along a dual special Lagrangian tori fibration on a mirror CY manifold which
interchanges the symplectic geometry and the complex geometry of a mirror pair.

In this paper we will explore the gauge theory formulation of six-dimensional Riemannian manifolds
to address the issue why CY manifolds exist with a mirror pair.
In order to simplify an underlying argumentation,
we will focus on orientable six-dimensional manifolds with spin structure.
In general relativity, the Lorentz group appears as the structure group acting on
orthonormal frames of the tangent bundle of a Riemannian manifold \ct{big-book}.
On the frame bundle, a Riemannian metric on spacetime manifold $M$ is replaced
by a local orthonormal basis $E_A \; (A = 1, \cdots, d)$ of the tangent bundle $TM$.
Then Einstein gravity can be formulated as a gauge theory of Lorentz group where
spin connections play a role of gauge fields and Riemann curvature tensors correspond
to their field strengths. On a six-dimensional Riemannian manifold $M$, for example,
local Lorentz transformations are orthogonal rotations in $Spin(6)$,
and spin connections $\omega_{AB} = \omega_{MAB} dx^M$ are the $spin(6)$-valued gauge fields
from the gauge theory point of view.\footnote{We will use large letters to indicate
a Lie group $G$ and small letters for its Lie algebra $\mathfrak{g}$.}
Then the Riemann curvature tensor $R_{AB} = d\omega_{AB} + \omega_{AC} \wedge \omega_{CB}$
precisely corresponds to the field strength of gauge fields $\omega_{AB}$
in $Spin(6)$ gauge theory. Since the Lie group $Spin(6)$ is isomorphic to $SU(4)$,
the six-dimensional Euclidean gravity can be formulated as an $SU(4)$ Yang-Mills gauge theory.
Via the gauge theory formulation of six-dimensional Riemannian manifolds,
we want to identify gauge theory objects corresponding to CY manifolds
and address their mirror symmetry from the perspective of Yang-Mills gauge theory.
To understand why there exists a mirror pair of CY manifolds, in particular,
we will employ the following well-known propositions for a $d$-dimensional Riemannian manifold $M$:

(A) The Riemann curvature tensors $R_{AB}$ are $spin(d)$-valued two-forms in $\Omega^2 (M) = \Lambda^2 T^*M$.

(B) There exists a global isomorphism between $d$-dimensional Lorentz groups and
classical Lie groups:
\begin{equation}\label{lie-iso}
\begin{array}{ll}
  Spin(3) \cong SU(2), \quad & Spin(4) \cong SU(2)_L \times SU(2)_R, \\
  Spin(5) \cong Sp(4), \quad & Spin(6) \cong SU(4).
\end{array}
\end{equation}

(C) There is an isomorphism between the Clifford algebra $\mathbb{C}l(d)$
in $d$-dimensions and the exterior algebra $\Lambda^* M$ of cotangent bundle $T^*M$
over $M$ \ct{spin-book,meinren}:\footnote{The space of the Clifford
algebra $\mathbb{C}l(d)$ is isomorphic, as a vector space, to the vector space of
the exterior algebra $\Lambda^* M$. This is not, however, an isomorphism of associative algebras
because the product in $\Lambda^* M$ is anticommutative while that in $\mathbb{C}l(d)$ is not
due to the central term in the Dirac algebra (\ref{dirac}).}
\begin{equation}\label{clifford-ext}
    \mathbb{C}l(d) \cong \Lambda^* M = \bigoplus_{k=0}^d \Omega^k (M)
\end{equation}
where $\Omega^k (M) = \Lambda^k T^* M$.

For the isomorphism (C) between the vector spaces, the ``volume operator" $\Gamma_{d+1} \equiv
\pm i^{\frac{d(d-1)}{2}} \Gamma^1 \cdots \Gamma^d$ in the Clifford algebra $\mathbb{C}l(d)$
corresponds to the Hodge-dual operator $*: \Omega^k (M) \to \Omega^{d-k} (M)$
in the exterior algebra $\Lambda^* M$
where $\Gamma^A \; (A=1, \cdots, d)$ are $d$-dimensional Dirac matrices obeying the Dirac algebra
\begin{equation}\label{dirac}
    \{ \Gamma^A, \Gamma^B \} = 2 \delta^{AB} \mathbf{I}_{2^{[\frac{d}{2}]}}.
\end{equation}
It is amusing to note that the Clifford algebra from a modern viewpoint can be thought of
as a quantization of the exterior algebra \cite{meinren}, in the same way that the Weyl algebra
is a quantization of the symmetric algebra.
In particular, the Clifford map \eq{clifford-ext} implies
that the Lorentz generators $J^{AB} \equiv \frac{1}{4}[\Gamma^A, \Gamma^B]$ in $\mathbb{C}l(d)$
are in one-to-one correspondence with two-forms in the space $\Omega^2 (M)$.
And the representation space of the Clifford algebra is a spinor vector space whose elements are called fermions
and essential ingredients in Standard Model. It may also be worthwhile to remark that any physical force is
represented by two-forms in the exterior algebra taking values in a classical Lie algebra.
In addition recall that the representation of
Clifford algebra in even dimensions is reducible and its irreducible representations are given by
chiral fermions. Then the isomorphism (C) implies that there must be a corresponding irreducible
decomposition of two-forms in $\Lambda^* M$. This fact, in our case, has a nontrivial consequence
for the Riemann curvature tensors $R_{AB} = \frac{1}{2} R_{ABCD} e^C \wedge e^D$ since
the $spin(d)$ Lie algebra indices $(A, B)$ and the form indices $(C, D)$ must have an identical structure
in a representation space of the Lorentz symmetry according to the isomorphism (C).
Our principal concern is then to pin down a geometrical consequence of the rudimentary fact (A)
after implementing the isomorphisms (B) and (C) to six-dimensional CY manifolds.

Let us briefly state the result summarized in the Table \ref{table-ms} in advance.
Compared to the four-dimensional case \cite{opy,ohya,loy,pos}, some acute changes arise.
First of all, there are two sources of two-forms on an orientable six-dimensional manifold $M$.
One is of course usual two-forms in $\Omega^2 (M)$ and the other is the Hodge-dual
of four-forms in $\Omega^4 (M)$. Therefore the vector space of two-forms is doubled in six dimensions:
\begin{equation}\label{doubling-6vec}
     \Lambda^2(M) \equiv \Omega^2 (M) \oplus * \Omega^4(M).
\end{equation}
The doubling of two-forms is resonant with the fact that
the irreducible representation of Lorentz symmetry is given by the chiral Lorentz generators
$J^{AB}_{\pm} \equiv \frac{1}{2} (\mathbf{I}_8 \pm \Gamma_7) J^{AB}$.
Definitely it corresponds to the mixture of two-forms
and four-forms in $\Lambda^2 (M)$ according to the correspondence $(\Gamma_7 \leftrightarrow *)$.
Since we need to take an irreducible representation of Lorentz symmetry, this demands us to think of
the irreducible components of Riemann curvature tensors as a sum of the usual curvature
tensors $R_{AB}$ and dual curvature tensors
defined by $\widetilde{R}_{AB} \equiv (*G)_{AB} = d {\widetilde{\omega}}_{AB}
+ {\widetilde{\omega}}_{AC} \wedge \widetilde{\omega}_{CB}$ where $G_{AB}$ is a 4-form tensor
taking values in $spin(6) \cong su(4)$ Lie algebra \cite{pos}.
Moreover it is necessary to impose the torsion-free condition for both spin connections, $\omega_{AB}$ and $\widetilde{\omega}_{AB}$, which leads to the symmetry property of the curvature tensors;
$R_{CDAB} = R_{ABCD}$ and $\widetilde{R}_{CDAB} = \widetilde{R}_{ABCD}$.
This is another reason why two kinds of indices $\big( [AB], [CD] \big)$ must be treated symmetrically
although they belong to different vector spaces.
To summarize, the Hodge duality admits two independent types of
curvature tensors $(R_{ABCD} \oplus \widetilde{R}_{ABCD})$ and they have to be
decomposed according to the irreducible representation of $spin(6) \cong su(4)$ Lie algebra.
In the end, the duplication of curvature tensors leads to the doubling for the variety
of six-dimensional spin manifolds.

It might be stressed that the doubling of six-dimensional spin manifolds is an inevitable
consequence of the elementary facts (A,B,C). It should be instructive to apply
the foregoing propositions (A,B,C) to four-manifolds to grasp their significance \cite{ohya,loy,pos}
although the four-dimensional situation is in stark contrast to the six-dimensional case.
In four dimensions, the Lorentz group $Spin(4)$ is isomorphic to $SU(2)_L \times SU(2)_R$
whose Lie algebras $su(2)_{L,R}$ consist of chiral Lorentz generators $J^{AB}_{\pm}
\equiv \Gamma_\pm J^{AB}$ with $\Gamma_\pm =  \frac{1}{2} (\mathbf{I}_4 \pm \Gamma_5)$
for chiral and anti-chiral representations.
The splitting of the Lie algebra, $spin(4) \cong su(2)_L \oplus su(2)_R$, is precisely
isomorphic to the canonical decomposition of the vector space $\Omega^2 (M)$ of two-forms:
\begin{equation}\label{2-form-dec}
    \Omega^2 (M) = \Omega_+^2(M) \oplus \Omega_-^2(M)
\end{equation}
where $\Omega_\pm^2 (M) \equiv P_\pm \Omega^2 (M)$ and $P_\pm = \frac{1}{2} (1 \pm *)$.
That is, the six-dimensional vector space $\Omega^2 (M)$ of two-forms splits canonically into
the sum of three-dimensional vector spaces of self-dual and anti-self-dual two-forms.
One can apply the canonical splitting of the two vector spaces to Riemann curvature tensors
simultaneously according to Eq. \eq{clifford-ext}.
It results in the well-known decomposition of the curvature tensor $R$ into irreducible
components \cite{ahs-riemann,besse}, schematically given by
\begin{equation}\label{dec-riemann}
    R = \left(
          \begin{array}{cc}
            W^+ + \frac{1}{12} s & B \\
            B^T & W^- + \frac{1}{12} s \\
          \end{array}
        \right)
\end{equation}
where $s$ is the scalar curvature, $B$ is the traceless Ricci tensor, and $W^\pm$ are the
(anti-)self-dual Weyl tensors. An important lesson from the four-dimensional example is that
the irreducible (chiral) representation of Lorentz symmetry corresponds to the canonical
split (\ref{2-form-dec}) of two-forms with the projection operator $P_\pm = \frac{1}{2} (1 \pm *)$.
We observe that the same analysis in six dimensions brings about a more dramatic result
due to the fact $6 = 2 + 4$. The doubling of six-dimensional spin manifolds will be
important to understand why CY manifolds arise with a mirror pair.

The gauge theory formulation of six-dimensional spin manifolds also leads to a valuable perspective
for the doubling. The first useful access is to identify a gauge theory object corresponding
to a CY 3-fold in the same sense that a gravitational instanton (or a hyper-K\"ahler manifold)
can be identified with an $SU(2)$ Yang-Mills instanton in four dimensions \cite{opy,ohya,loy}.
An obvious guess goes toward a six-dimensional generalization of the four-dimensional Yang-Mills
instantons known as Hermitian Yang-Mills (HYM) instantons.
Indeed this relationship has been well-known to string theorists and mathematicians under the name
of the Donaldson-Uhlenbeck-Yau (DUY) theorem \cite{duy}.
We quote a paragraph in \ct{cy-book} (211 page) to clearly summarize this picture.

\begin{quote}
The point of intersection between the Calabi conjecture and the DUY theorem
is the tangent bundle. And here's why: Once you've proved the existence of
CY manifolds, you have not only those manifolds but their tangent bundles
as well, because every manifold has one. Since the tangent bundle is defined
by the CY manifold, it inherits its metric from the parent manifold (in
this case, the CY). The metric for the tangent bundle, in other words,
must satisfy the CY equations. It turns out, however, that for the tangent
bundle, the Hermitian Yang-Mills equations are the same as the CY equations,
provided the background metric you've selected is the CY. Consequently,
the tangent bundle, by virtue of satisfying the CY equations,
automatically satisfies the Hermitian Yang-Mills equations, too.
\end{quote}

If a CY manifold $M$ can be related to a HYM instanton,
a natural question immediately arises. Since a CY manifold $M$ has a mirror manifold,
there will be a mirror CY manifold $\widetilde{M}$ obeying the mirror relation \eq{mirror-cyh1}.
This in turn implies that there must be a {\it mirror} HYM instanton
which can be derived from the mirror CY manifold $\widetilde{M}$.
Thus we want to understand the relation between the HYM instanton and its mirror instanton
from the gauge theory perspective. Since the Lorentz group $Spin(6)$ is isomorphic to $SU(4)$,
the chiral and anti-chiral representations $\mathbf{4}$ and $\mathbf{4}'$ of $Spin(6)$
are equivalent to the fundamental and anti-fundamental representations $\mathbf{4}$
and $\overline{\mathbf{4}}$ of $SU(4)$. Recall that the fundamental representation $\mathbf{4}$
of $SU(4)$ is a complex representation and so its complex conjugate $\overline{\mathbf{4}}$
is an inequivalent representation different from $\mathbf{4}$.
Therefore, given a CY manifold $M$, one can embed the HYM instanton inherited from $M$
into two different representations. But this situation is equally true for the mirror
CY manifold $\widetilde{M}$. Thus there is a similar doubling for the variety of
HYM instantons as occurred to CY manifolds, as summarized in the Table 1.

It may be interesting to compare this situation with the four-dimensional case \cite{loy,pos}.
In four dimensions, the positive and negative chirality spinors of $Spin(4)$ are
given by $SU(2)_L$ and $SU(2)_R$ spinors, $\mathbf{2}_L$ and $\mathbf{2}_R$, respectively.
In this case, it is necessary to have two independent $SU(2)$ factors to be compatible
with the splitting (\ref{2-form-dec})
because the irreducible representation of $SU(2)$ is real. It is interesting to see how
(A,B,C) take part in the conspiracy. First, a CY 2-fold can be mapped to a self-dual or $SU(2)_L$ instanton
which lives in the chiral representation $\mathbf{2}_L$, while a mirror CY 2-fold is isomorphically
related to an anti-self-dual or $SU(2)_R$ instanton in the anti-chiral representation $\mathbf{2}_R$.
For this correspondence, the $SU(2)$ gauge group of Yang-Mills instantons is identified with
the holonomy group of CY 2-folds. This picture is generalized to six dimensions in an interesting way.
In six dimensions, the canonical splitting (\ref{2-form-dec}) is applied
to the enlarged vector space \eq{doubling-6vec} as
\begin{equation}\label{split-6d}
    \Lambda^2(M) = \Omega^2_+ (M) \oplus \Omega^2_-(M)
\end{equation}
where the decomposition $\Omega^2_\pm (M)$ is dictated by the chiral splitting
$J^{AB} = J^{AB}_+ \oplus J^{AB}_-$ according to the isomorphism (C).
From the gauge theory perspective, the splitting (\ref{split-6d}) is also compatible with
the fundamental and anti-fundamental representations of the gauge group $SU(4) \cong Spin(6)$
because the chiral representation of $Spin(6)$ is identified with the fundamental
representation of $SU(4)$. After all, we will get the picture that
the HYM instanton on $TM$ embedded in the fundamental
representation $\mathbf{4}$ is mirror to the HYM instanton
on $T\widetilde{M}$ in the anti-fundamental representation $\overline{\mathbf{4}}$.
This structure is summed up in the Table 1, where
$CY3$ refers to a CY 3-fold $M$ and $\widetilde{CY3}$ its mirror $\widetilde{M}$.
And $HYM$ denotes a HYM instanton on $M$ in the complex
representation either $\mathbf{3}$ or $\overline{\mathbf{3}}$ of $SU(3) \subset SU(4)$
and $\widetilde{HYM}$ its mirror on $\widetilde{M}$ in the opposite complex representation.

\begin{table}
  \centering
  \begin{tabular}{|c|c|c|}
    \hline
    $M$ & $Spin(6)$ & $SU(4)$ \\
    \hline
    $\mathbb{A}: \mathbf{3} \hookrightarrow \mathbf{4}$ & $CY3$ & $HYM$ \\
    \hline
    $\mathbb{B}: \overline{\mathbf{3}} \hookrightarrow \overline{\mathbf{4}}$ & $CY3$ & $HYM$ \\
    \hline
  \end{tabular}
  \begin{tabular}{c}
  ${}$ \\
  $\Longleftrightarrow$ \\
  $\Longleftrightarrow$
\end{tabular}
  \begin{tabular}{|c|c|c|}
    \hline
   $\widetilde{M}$ & $Spin(6)$ & $SU(4)$ \\
    \hline
   $\mathbb{B}: \overline{\mathbf{3}} \hookrightarrow \overline{\mathbf{4}}$ & $\widetilde{CY3}$ & $\widetilde{HYM}$ \\
    \hline
   $\mathbb{A}: \mathbf{3} \hookrightarrow \mathbf{4}$ &  $\widetilde{CY3}$ & $\widetilde{HYM}$ \\
    \hline
  \end{tabular}
\caption{Mirror symmetry}\label{table-ms}
\end{table}

The purpose of this paper is to understand the structure in the Table \ref{table-ms}.
Up to our best knowledge, there is no concrete work to address the mirror symmetry
based on the picture in the Table \ref{table-ms}
although the mirror symmetry has been extensively studied so far.
We will show that CY manifolds and HYM instantons exist with mirror pairs
as a consequence of the doubling (\ref{doubling-6vec}) of two-forms in six dimensions.
It is arguably a remarkable consequence of the mysterious Clifford isomorphism (C).

This paper is organized as follows. In section 2, we formulate $d$-dimensional
Euclidean gravity as a $Spin(d)$ Yang-Mills gauge theory. The explicit relations between
gravity and gauge theory variables are established. In particular, we construct
the dual curvature tensors $\widetilde{R}_{AB} \equiv (*G)_{AB} = d {\widetilde{\omega}}_{AB}
+ {\widetilde{\omega}}_{AC} \wedge \widetilde{\omega}_{CB}$ that are necessary
for an irreducible representation of Lorentz symmetry. We observe that the geometric structure
described by dual spin connections ${\widetilde{\omega}}_{AB}$ and curvature tensors $\widetilde{R}_{AB}$
is exactly parallel to the usual one described by $(\omega_{AB}, R_{AB})$ and so clarify
why the variety of orientable spin manifolds is doubled.

We apply in section 3 the gauge theory formulation to six-dimensional
Riemannian manifolds. For that purpose we devise a six-dimensional version
of the 't Hooft symbols which realizes the isomorphism between $spin(6)$ Lorentz algebra
and $su(4)$ Lie algebra. As the $spin(6)$ Lorentz algebra has two
irreducible spinor representations, there are accordingly two kinds of the 't Hooft symbols
depending on the chirality of irreducible $spin(6)$ representations.
Our construction of six-dimensional 't Hooft symbols is new up to our best knowledge.
Using this construction, we impose the K\"ahler condition on the 't Hooft symbols.
This is done by projecting the 't Hooft symbols to $U(3)$-valued ones and
so results in the reduction of the gauge group from $SU(4)$ to $U(3)$.
After imposing the Ricci-flat condition, the gauge group in Yang-Mills gauge theory is
further reduced to $SU(3)$. This result is utilized to show that six-dimensional CY manifolds can be
recast as HYM instantons in $SU(3)$ Yang-Mills gauge theory. We elucidate why the canonical
splitting \eq{split-6d} of six-dimensional spin manifolds corresponds to the chiral representation
of $Spin (6)$. It turns out that this splitting is equally applied to CY manifolds as well
as HYM instantons.

In section 4, we apply the results in section 3 to CY manifolds to see how the mirror symmetry
between them can be explained by the doubling of six-dimensional spin manifolds.
We observe that it is always possible to find a pair of CY manifolds such that their Euler characteristics
in different chiral representations obey the mirror relation (\ref{mirror-cyh1}).
This implies that a pair of CY manifolds in the opposite chiral representation are mirror to each other
as indicated by the arrow $(\Longleftrightarrow)$ in the Table 1.

In section 5, we revisit the relation between CY manifolds and HYM instantons to
discuss the mirror symmetry from a completely gauge theory perspective.
We show that a pair of HYM instantons embedded in different complex representations
$\mathbf{4}$ and $\overline{\mathbf{4}}$ correspond to a mirror pair of CY manifolds
as summarized in the Table \ref{table-ms}. This result is consistent with the mirror symmetry
because the integral of the third Chern class $c_3(E)$ for a vector bundle $E$ is equal to
the Euler characteristic of tangent bundle $TM$ when $E=TM$ and
the third Chern class has a desired sign flip between a complex vector bundle $E$ in the fundamental
representation and its conjugate bundle $\overline{E}$ in the anti-fundamental representation.
Therefore we confirm the picture in the Table \ref{table-ms} that the mirror symmetry between CY manifolds
can be understood as a mirror pair of HYM instantons
in holomorphic vector bundles.

Finally we recapitulate in section 6 the results obtained in this paper
and conclude the paper with a few speculative remarks.

In appendix A, we fix the basis for the chiral representation of $Spin(6)$ and
the fundamental representation of $SU(4)$ and list their structure constants.
In appendix B, we present an explicit construction of the six-dimensional 't Hooft symbols
and their algebraic properties in each chiral basis.

\section{Gravity As A Gauge Theory}

In this section we consider the gauge theory formulation of Riemannian manifolds taking values
in an irreducible spinor representation of the Lorentz group \cite{loy,pos}.
This section is to establish the notation for the doubled variety
of Riemannian manifolds, but more detailed exposition will be deferred to the next section.
On a Riemannian manifold $M$ of dimension $d$, the spin connection is
a $spin(d)$-valued one-form and can be identified, in general, with a $Spin(d)$ gauge field.
In order to make an explicit identification between the spin connections and the
corresponding gauge fields, let us first consider the $d$-dimensional Dirac algebra (\ref{dirac})
where $\Gamma^A \; (A = 1, \cdots, d)$ are Dirac matrices. Then the $spin(d)$ Lorentz generators
are given by
\begin{equation}\label{lorentz-gen}
   J^{AB} = \frac{1}{4} [\Gamma^A, \Gamma^B]
\end{equation}
which satisfy the following Lorentz algebra
\begin{equation}\label{lorentz-algebra}
    [J^{AB}, J^{CD}] = - \big(\delta^{AC} J^{BD} - \delta^{AD}J^{BC}
    - \delta^{BC}J^{AD} + \delta^{BD}J^{AC} \big).
\end{equation}
The spin connection is defined by $\varpi = \frac{1}{2} \varpi_{AB} J^{AB}$, which transforms
in the standard way as a $Spin(d)$ gauge field under local Lorentz transformations
\begin{equation}\label{spin-lorentz-tr}
    \varpi \; \to \; \varpi' = \Lambda \varpi \Lambda^{-1} + \Lambda d \Lambda^{-1}
\end{equation}
where $\Lambda = e^{\frac{1}{2}\lambda_{AB}(x) J^{AB}} \in Spin(d)$.

In even dimensions, the spinor representation is reducible and its irreducible representations
are given by positive and negative chiral representations. In next section we will provide an explicit
chiral representation for the six-dimensional case. The Lorentz generators
for the chiral representation are given by
\begin{equation}\label{lorents-chiral}
    J^{AB} = \left(
               \begin{array}{cc}
                 J^{AB}_+ & 0 \\
                 0 & J^{AB}_- \\
               \end{array}
             \right)
\end{equation}
where $J^{AB}_\pm = \Gamma_\pm J^{AB}$ and $\Gamma_\pm = \frac{1}{2} \big( \mathbf{I}_{2^{[\frac{d}{2}]}}
\pm \Gamma_{d+1} \big)$.
Therefore the spin connection in the chiral representation takes the form
\begin{equation}\label{spin-chiral}
\varpi = \frac{1}{2} \varpi_{AB} J^{AB} = \left(
               \begin{array}{cc}
                 \omega^{(+)} & 0 \\
                 0 & \omega^{(-)} \\
               \end{array}
             \right) = \frac{1}{2} \left(
               \begin{array}{cc}
                 \omega^{(+)}_{AB} J^{AB}_+ & 0 \\
                 0 & \omega^{(-)}_{AB} J^{AB}_- \\
               \end{array}
             \right).
\end{equation}
Here we used a sloppy notation for $\varpi$ which must be understood as $\varpi =
\frac{1}{2} \varpi_{\mathbf{A}\mathbf{B}} J^{\mathbf{A}\mathbf{B}}$ where $\mathbf{A} = (A, A'),
\; \mathbf{B} = (B, B')$ and $\varpi_{AB'} = \varpi_{A'B} = 0$.
For a notational simplicity we will use this notation since it will not introduce too much confusion.
Note that the spin connections $\omega^{(+)}$ and $\omega^{(-)}$ are considered as
independent since they are resulted from the doubling of one-forms due to the Hodge duality,
as will be shown later.

Now we introduce a $Spin(d)$ gauge field defined by
\begin{equation}\label{gauge-chiral}
\mathbb{A} = A^a \mathbb{T}^a = \left(
               \begin{array}{cc}
                 A^{(+)a}T^a & 0 \\
                 0 & A^{(-)a} (T^a)^* \\
               \end{array}
             \right), \qquad \mathbb{T}^a =  \left(
               \begin{array}{cc}
                 T^a & 0 \\
                 0 & (T^a)^* \\
               \end{array}
             \right)   \in spin(d),
\end{equation}
where $A^{(\pm)a} = A^{(\pm)a}_{M} dx^M \; (a = 1, \cdots, \frac{d(d-1)}{2})$
are one-form connections on $M$. We will take the definition \eq{gauge-chiral} by adopting the group
isomorphism (\ref{lie-iso}). The Lie algebra generators are matrices obeying the commutation relation
\begin{equation}\label{lie-algebra}
    [\mathbb{T}^a, \mathbb{T}^b] = - f^{abc} \mathbb{T}^c
\end{equation}
where $T^a$ and $(T^a)^*$ are generators in a representation $R$ and its conjugate representation $\overline{R}$, respectively. The identification we want to make is then given by
\begin{equation}\label{id}
    \varpi = \frac{1}{2} \varpi_{AB} J^{AB} \cong \mathbb{A} = A^a \mathbb{T}^a.
\end{equation}
Then the Lorentz transformation \eq{spin-lorentz-tr} can be interpreted as
a usual gauge transformation
\begin{equation}\label{gauge-lorentz-tr}
    \mathbb{A} \; \to \; \mathbb{A}' = \Lambda \mathbb{A} \Lambda^{-1} + \Lambda d \Lambda^{-1}
\end{equation}
where $\Lambda = e^{\lambda^a(x) \mathbb{T}^a} \in Spin(d)$.
The Riemann curvature tensor is defined by  \ct{big-book}
\begin{eqnarray} \la{riemann-tensor}
  \mathcal{R} &=&  \frac{1}{2} \mathcal{R}_{AB} J^{AB} = d \varpi + \varpi \wedge \varpi = \frac{1}{2} \Big(d \varpi_{AB}
  + \varpi_{AC} \wedge \varpi_{CB} \Big) J^{AB} \xx
  &=& \frac{1}{2} \left(
               \begin{array}{cc}
                 R^{(+)}_{AB} J^{AB}_+ & 0 \\
                 0 & R^{(-)}_{AB} J^{AB}_- \\
               \end{array}
             \right),
\end{eqnarray}
where $R^{(\pm)}_{AB} = \frac{1}{2} \Big( \partial_M \omega^{(\pm)}_{NAB} - \partial_N \omega^{(\pm)}_{MAB}
  + \omega^{(\pm)}_{MAC} \omega^{(\pm)}_{NCB} -  \omega^{(\pm)}_{NAC} \omega^{(\pm)}_{MCB}
  \Big) dx^M \wedge dx^N$. Or, in terms of gauge theory variables, it is given by
\begin{eqnarray} \la{curvature-tensor}
  \mathbb{F} &=& F^a \mathbb{T}^a = d \mathbb{A} + \mathbb{A} \wedge \mathbb{A}
  = \Big(d A^a - \frac{1}{2} {f_{bc}}^a A^b \wedge A^c \Big) \mathbb{T}^a \xx
  &=& \frac{1}{2} \left(
               \begin{array}{cc}
                 F^{(+)a} T^a & 0 \\
                 0 & F^{(-)a} (T^a)^* \\
               \end{array}
             \right),
\end{eqnarray}
where $F^{(\pm)a} = \frac{1}{2} \Big(\partial_M A^{(\pm)a}_{N} - \partial_N A^{(\pm)a}_{M}
  - f^{abc} A^{(\pm)b}_{M} A^{(\pm)c}_{N} \Big) dx^M \wedge dx^N$.

As we outlined in section 1, in addition to the usual curvature tensor $R_{AB} = d {\omega}_{AB}
+ \omega_{AC} \wedge \omega_{CB}$, we need to introduce the dual curvature tensor defined by
\begin{equation}\label{dual-curvature}
\widetilde{R}_{AB} \equiv (*G)_{AB} = d {\widetilde{\omega}}_{AB}
+ {\widetilde{\omega}}_{AC} \wedge \widetilde{\omega}_{CB}
\end{equation}
where $G_{AB}$ is a $(d-2)$-form tensor taking values in $spin(d)$ Lie algebra.
One may consider the dual spin connection ${\widetilde{\omega}}_{AB} \equiv (*\theta)_{AB}$ as
the Hodge-dual of a $(d-1)$-form $\theta_{AB}$ in $spin(d)$ Lie algebra.
It is useful to introduce the adjoint exterior differential operator $\delta : \Omega^k (M) \to
\Omega^{k-1} (M)$ defined by
\begin{equation}\label{adjoint-edo}
    \delta = (-1)^{dk + d + 1} * d *
\end{equation}
where the Hodge-dual operator $*: \Omega^k (M) \to \Omega^{d-k} (M)$ obeys
the well-known relation
\begin{equation}\label{star-sq}
*^2 \alpha = (-1)^{k (d - k)} \alpha
\end{equation}
for $\alpha \in \Omega^k (M)$. Using the adjoint differential operator $\delta$,
the $spin(d)$-valued $(d-2)$-form $G_{AB}$ in Eq. (\ref{dual-curvature}) can be written as
\begin{equation}\label{4-form}
 G_{AB} = (-)^{d-1} \delta \theta_{AB} + \theta_{AC} \barwedge \theta_{CB}
\end{equation}
where we devised a simplifying notation
\begin{equation}\label{barwedge}
\alpha \barwedge \beta \equiv * \Bigl( (* \alpha) \wedge (* \beta) \Bigr) \in \Omega^{p+q-d} (M)
\end{equation}
for $\alpha \in \Omega^p (M)$ and $\beta \in \Omega^q (M)$. Using the nilpotency of the adjoint
differential operator $\delta$, i.e. $\delta^2 = 0$, one can derive the (second) Bianchi identity
\begin{equation}\label{dual2-bianchi}
    \delta G_{AB} + (-)^{d-1} \bigl( \theta_{AC} \barwedge G_{CB} - G_{AC} \barwedge \theta_{CB} \bigr) = 0.
\end{equation}
It may be compared with the ordinary Bianchi identity in general relativity written
as $$d R_{AB} + \omega_{AC} \wedge R_{CB} - R_{AC} \wedge \omega_{CB} = 0.$$

Let us also introduce dual vielbeins $\widetilde{e}^A \equiv (* h)^A$ where $h^A \in \Omega^{d-1} (M)$,
in addition to the usual vielbeins $e^A \; (A=1, \cdots, d)$
which independently form a local orthonormal coframe at each spacetime point in $M$.
We combine the dual one-forms $\widetilde{e}^A$ with the usual coframe $e^A$ to define
a matrix of vielbeins
\begin{equation}\label{e-matrix}
    \mathfrak{E} = \mathfrak{E}_A \Gamma^A = \left(
                   \begin{array}{cc}
                     0 & e^{(+)A} \gamma^A \\
                     e^{(-)A} \overline{\gamma}^A & 0 \\
                   \end{array}
                 \right)
\end{equation}
where
\begin{equation}\label{pair-vielbein}
    e^{(\pm)A} \equiv \frac{1}{2} (e^A \pm \widetilde{e}^A).
\end{equation}
The coframe basis $\{ e^{(\pm)A} \in \Gamma(T^* M) \}$ defines dual vectors $E^{(\pm)}_A
= E^{(\pm)M}_A \partial_M \in \Gamma(TM)$ by a natural pairing
\begin{equation} \label{dual-vector}
\langle e^{(\pm)A}, E^{(\pm)}_B \rangle = \delta^A_B.
\end{equation}
The above pairing leads to the relation $e^{(\pm)A}_M E^{(\pm)M}_B = \delta^A_B$.
Since we regard the spin connections $\omega^{(+)}$ and $\omega^{(-)}$ as independent,
let us consider two kinds of geometrical data on a spin manifold $M$,
dubbed $\mathbb{A}$ and $\mathbb{B}$ classes:
\begin{equation}\label{pair-data}
\begin{array}{l}
  \mathbb{A}: \big(e^{(+)A}, \omega^{(+)}_{AB} \big), \\
  \mathbb{B}: \big(e^{(-)A}, \omega^{(-)}_{AB} \big).
\end{array}
\end{equation}
We emphasize that the geometric structure of a $d$-dimensional spin manifold can be described by
either the type $\mathbb{A}$ or the type $\mathbb{B}$ but they should be regarded as independent
even topologically. In other words, we can separately consider a Riemannian metric for each class given by
\begin{eqnarray} \label{6-metric}
ds^2_\pm &=& \delta_{AB} e^{(\pm)A} \otimes e^{(\pm)B} = \delta_{AB} e^{(\pm)A}_M e^{(\pm)B}_N
\; dx^M \otimes dx^N \nonumber
\\ &\equiv& g^{(\pm)}_{MN}(x) \; dx^M \otimes dx^N
\end{eqnarray}
or
\begin{eqnarray} \label{inverse-metric}
\Bigl(\frac{\partial}{\partial s}\Bigr)^2_\pm &=& \delta^{AB} E^{(\pm)}_A \otimes E^{(\pm)}_B
= \delta^{AB} E^{(\pm)M}_A E^{(\pm)N}_B \; \partial_M \otimes \partial_N \nonumber
\\ &\equiv& g_{(\pm)}^{MN}(x)
\; \partial_M \otimes \partial_N.
\end{eqnarray}

In order to recover general relativity from the gauge theory formulation of gravity,
it is necessary to impose the torsion-free condition, i.e.,
\begin{equation}\label{double-torsion}
    T^{(\pm)A} = d e^{(\pm)A} + {\omega^{(\pm)A}}_{B} \wedge e^{(\pm)B} = 0.
\end{equation}
As a result, the spin connections are determined by vielbeins, i.e. $\omega^{(\pm)}
= \omega^{(\pm)} \big( e^{(\pm)} \big)$, from which
one can deduce the first Bianchi identity
\begin{equation}\label{2-1bianchi}
 R^{(\pm)}_{AB} \wedge e^{(\pm)B} = 0
\end{equation}
where the curvature tensors $R^{(\pm)}_{AB}$ are defined by Eq. (\ref{riemann-tensor}).
It is not difficult to see that Eq. (\ref{2-1bianchi}) leads to the symmetry property
for the Riemann curvature tensors
$R^{(\pm)}_{AB} \equiv \frac{1}{2} R^{(\pm)}_{ABCD} e^{(\pm)C} \wedge e^{(\pm)D}$;
$R^{(\pm)}_{ABCD} = R^{(\pm)}_{CDAB}$. It may be convenient to introduce the torsion matrix $\mathbb{T}$
defined by
\begin{eqnarray}\label{torsion-matrix}
    \mathbb{T} &=& d\mathfrak{E} + \varpi \wedge \mathfrak{E} + \mathfrak{E} \wedge \widehat{\varpi} \xx
               &=& \left(
                   \begin{array}{cc}
                     0 & T^{(+)A} \gamma^A \\
                     T^{(-)A} \overline{\gamma}^A & 0 \\
                   \end{array}
                 \right),
\end{eqnarray}
where we have defined the inverted spin connection $\widehat{\varpi} \equiv
\frac{1}{2} \widehat{\varpi}_{AB} \widetilde{J}^{AB} = \frac{1}{2} \left(
              \begin{array}{cc}
                \omega_{AB}^{(-)} J^{AB}_- & 0 \\
                 0 & \omega_{AB}^{(+)} J^{AB}_+ \\
               \end{array}
             \right)$.
It is straightforward to show that
\begin{equation}\label{dt-matrix}
    d \mathbb{T} = \frac{1}{2} \left(
                                 \begin{array}{cc}
                                   0 & R^{(+)}_{AB} \wedge e^{(+)B} \gamma^A \\
                                   R^{(-)}_{AB} \wedge e^{(-)B} \overline{\gamma}^A & 0 \\
                                 \end{array}
                               \right)
\end{equation}
and so the first Bianchi identity (\ref{2-1bianchi}) is automatic because of
the torsion-free condition, $\mathbb{T} = 0$.
Similarly, using the definition (\ref{riemann-tensor}), it is easy to derive the second Bianchi
identity, $\mathcal{D} \mathcal{R} \equiv d\mathcal{R} + \varpi \wedge \mathcal{R}
- \mathcal{R} \wedge \varpi = 0$, whose matrix form reads as
\begin{equation}\label{dr-matrix}
\mathcal{D} \mathcal{R} = \left(
                            \begin{array}{cc}
                              D^{(+)} R^{(+)} & 0 \\
                              0 & D^{(-)} R^{(-)} \\
                            \end{array}
                          \right) = 0
\end{equation}
where
\begin{equation}\label{2+--bianchi}
 D^{(\pm)} R^{(\pm)} \equiv d R^{(\pm)} + \omega^{(\pm)} \wedge R^{(\pm)} - R^{(\pm)} \wedge \omega^{(\pm)}.
\end{equation}
In terms of gauge theory variables, it can be stated as $\mathbb{D} \mathbb{F} \equiv
d \mathbb{F} + \mathbb{A} \wedge \mathbb{F} - \mathbb{F} \wedge \mathbb{A} = 0$ or
$D^{(\pm)} F^{(\pm)} \equiv d F^{(\pm)} + A^{(\pm)} \wedge F^{(\pm)} - F^{(\pm)} \wedge A^{(\pm)} = 0$.

To sum up, a $d$-dimensional orientable Riemannian manifold admits a globally defined volume form which
leads to the isomorphism between $\Omega^k (M)$ and $\Omega^{d-k} (M)$.
In particular, it doubles the two-form vector space which leads to the enlargement for the geometric
structure of Riemannian manifolds. The Hodge duality $*: \Omega^k (M) \to \Omega^{d-k} (M)$ is thus
the origin of the doubling for the variety of Riemannian manifolds. One is described by $(e^A, \omega_{AB}, R_{AB})$
and the other independent construction is given by $(\widetilde{e}^A, \widetilde{\omega}_{AB}, \widetilde{R}_{AB})
\cong * (h^A, \theta_{AB}, G_{AB})$. According to the isomorphism (C), they are decomposed into
two irreducible representations of Lorentz symmetry.
In next section we will apply the irreducible decomposition to six-dimensional spin manifolds
to see why the variety of Riemannian manifolds is doubled.

\section{Spinor Representation of Six-dimensional Riemannian Manifolds}

We will apply the gauge theory formulation in the previous section to
six-dimensional Riemannian manifolds. For this purpose, the $Spin(6)$ Lorentz group for Euclidean gravity
will be identified with the $SU(4)$ gauge group in Yang-Mills gauge theory.
A motivation for the gauge theory formulation of six-dimensional Euclidean gravity is
to identify a gauge theory object corresponding to a CY manifold and
to understand the mirror symmetry of CY manifolds in terms of Yang-Mills gauge theory.
Because our gauge theory formulation is based on the identification \eq{id},
we will restrict ourselves to orientable six-dimensional manifolds with spin structure and
consider a spinor representation of $Spin(6)$ in order to scrutinize the relationship.

Let us start with the Clifford algebra $\mathbb{C}l(6)$ whose generators are given by
\begin{equation}\label{clifford-gen}
    \mathbb{C}l(6) = \{\mathbf{I}_8, \Gamma^A, \Gamma^{AB}, \Gamma^{ABC},
     \Gamma_7 \Gamma^{AB}, \Gamma_7 \Gamma^A, \Gamma_7 \}
\end{equation}
where $\Gamma^A \;(A=1, \cdots, 6)$ are six-dimensional Dirac matrices satisfying
the algebra \eq{dirac} and $\Gamma^{A_1A_2 \cdots A_k} \equiv \frac{1}{k!}
\Gamma^{[A_1} \Gamma^{A_2} \cdots \Gamma^{A_k]}$ assumes the complete antisymmetrization
of indices. $\Gamma_7 \equiv i \Gamma^1 \cdots \Gamma^6$ is the chiral matrix given by \eq{gamma7}.
According to the isomorphism (\ref{clifford-ext}), the Clifford algebra \eq{clifford-gen}
can be isomorphically mapped to the exterior algebra of a cotangent bundle $T^* M$
\begin{equation}\label{cliff-ext}
   \mathbb{C}l(6) \cong \Lambda^* M = \bigoplus_{k=0}^6 \Omega^{k} (M)
\end{equation}
where the chirality operator $\Gamma_7$ corresponds to the Hodge dual operator $* : \Omega^{k} (M)
\to \Omega^{6-k} (M)$.

The spinor representation of $Spin(6)$ can be constructed by 3 fermion creation
operators $a_i^* \;(i=1,2,3)$ and the corresponding annihilation
operators $a^j \;(j=1,2,3)$ (see appendix 5.A in \ct{string-book}).
This fermionic system can be represented in a Hilbert space $V$ of dimension 8
with a Fock vacuum $|\Omega\rangle$, annihilated by all the annihilation operators.
The states in $V$ are obtained by acting the product of $k$ creation
operators $a^*_{i_1} \cdots a^*_{i_k}$ on the vacuum $|\Omega\rangle$, i.e.,
\begin{equation}\label{spin-6}
     V = \bigoplus_{k=0}^3 |\Omega_{i_1 \cdots i_k} \rangle =
\bigoplus_{k=0}^3 a^*_{i_1} \cdots a^*_{i_k} |\Omega\rangle.
\end{equation}
The spinor representation of the algebra (\ref{clifford-gen}) is reducible and has two irreducible spinor representations. Indeed the Hilbert space $V$ splits into the spinors $S_\pm$ of positive and negative chirality,
i.e. $V = S_+ \oplus S_-$, each of dimension 4.
If the Fock vacuum $|\Omega\rangle$ has positive chirality,
the positive chirality spinors of $Spin(6)$ are states given by
\begin{equation}\label{+-spinor}
    S_+ = \bigoplus_{k \;\mathrm{even}} |\Omega_{i_1 \cdots i_k} \rangle =
    |\Omega\rangle + |\Omega_{ij} \rangle \equiv \mathbf{4}
\end{equation}
while the negative chirality spinors are those obtained by
\begin{equation}\label{--spinor}
    S_- = \bigoplus_{k \;\mathrm{odd}} |\Omega_{i_1 \cdots i_k} \rangle =
    |\Omega_i\rangle + |\Omega_{ijk} \rangle \equiv \overline{\mathbf{4}}.
\end{equation}
As $spin(6)$ Lorentz algebra is isomorphic to $su(4)$ Lie algebra,
the positive and negative chirality spinors of $spin(6)$ can be identified with the fundamental
representation $\mathbf{4}$ and the anti-fundamental representation $\overline{\mathbf{4}}$
of $su(4)$, respectively \ct{string-book}.
As a result, the chiral spinor representations $S_+$ and $S_-$ of $Spin(6)$ are identified
with the fundamental representations $\mathbf{4}$ and $\overline{\mathbf{4}}$ of $SU(4)$.

One can form a direct product of the fundamental representations $\mathbf{4}$
and $\overline{\mathbf{4}}$ in order to classify the Clifford generators
in Eq. \eq{clifford-gen}:
\begin{eqnarray} \la{44*-clifford}
&&  \mathbf{4} \otimes \overline{\mathbf{4}}  = \mathbf{1} \oplus \mathbf{15} =
\{\Gamma_+, \Gamma^{AB}_+ \},  \\
\la{4*4-clifford}
&&  \overline{\mathbf{4}} \otimes \mathbf{4}  = \mathbf{1} \oplus \mathbf{15} =
\{\Gamma_-, \Gamma^{AB}_- \}, \\
\la{44-clifford}
&&  \mathbf{4} \otimes \mathbf{4}  = \mathbf{6} \oplus \mathbf{10} =
\{ \Gamma^A_+, \Gamma_+^{ABC} \},  \\
\la{4*4*-clifford}
&&  \overline{\mathbf{4}} \otimes \overline{\mathbf{4}} = \mathbf{6} \oplus \mathbf{10}=
\{\Gamma^A_-, \Gamma_-^{ABC} \},
\end{eqnarray}
where $\Gamma_\pm \equiv \frac{1}{2}(\mathbf{I}_8 \pm \Gamma_7)$ are the projection operators onto
the space of definite chirality and $\Gamma_\pm^{A_1A_2 \cdots A_k}
\equiv \Gamma_\pm \Gamma^{A_1A_2 \cdots A_k}$.
Note that $\mathbf{15}$ in Eqs. \eq{44*-clifford} and \eq{4*4-clifford} is the
adjoint representation of $SU(4)$ and $\mathbf{6}$ and $\mathbf{10}$ in Eqs. \eq{44-clifford}
and \eq{4*4*-clifford} are the antisymmetric and symmetric representations of $SU(4)$, respectively.
See appendix A for the Lie algebra generators in the chiral representation of $Spin(6)$ and
the fundamental representation of $SU(4)$.
It is important to notice that $\Gamma_+^{AB} \in \mathbf{15}$ and $\Gamma_-^{AB} \in \mathbf{15}$
are independent of each other, i.e. $[\Gamma^{AB}_+, \Gamma^{CD}_-] = 0$,
and this doubling of the Clifford basis is parallel to the doubling of two-forms
according to the Clifford isomorphism (\ref{cliff-ext}) as will be clarified below.

We want to find the irreducible decomposition of Riemann curvature tensors
under the Lorentz symmetry as the six-dimensional version of Eq. \eq{dec-riemann}.
As we noticed before, there are two kinds of Lorentz generators given by
the irreducible components $\Gamma^{AB}_\pm = \Gamma_\pm \Gamma^{AB}$ which correspond to
the chiral and anti-chiral representations of Lorentz algebra $spin(6) \cong su(4)$.
Recall that Eq. \eq{riemann-tensor} takes the following split of curvature
tensors $\mathcal{R} = \frac{1}{2} \mathcal{R}_{AB} e^A \wedge e^B$:
\begin{equation}\label{id-f}
    \mathcal{R}_{AB} = R_{AB}^{(+)} \oplus R_{AB}^{(-)}
    = \bigl(F_{AB}^{(+)a} T_1^a \oplus F_{AB}^{(-)a} T^a_2 \bigr) = \mathbb{F}_{AB}
\end{equation}
where $R_{AB}^{(\pm)} = \Gamma_\pm \mathcal{R}_{AB}$ and both $T^a_1$ and $T^a_2$ obey the $su(4)$
Lie algebra defined by (\ref{lie-algebra}).
The doubling of $su(4)$ Lie algebra in four-dimensional representations $R_1$ and $R_2$
on the right-hand side was considered in parallel to the spinor representation on the left-hand side.
Since the Lorentz generators $J^{AB}_\pm = \Gamma_\pm J^{AB}$ are in one-to-one correspondence
with two-forms in the vector space (\ref{split-6d}), we identify the following map:
\begin{equation}\label{22-forms}
  J^{AB}_+ \quad \leftrightarrow \quad F^{(+)}_{AB}, \qquad
  J^{AB}_- \quad \leftrightarrow \quad F^{(-)}_{AB}.
\end{equation}
Since the role of the chiral operator $\Gamma_7$ is parallel with the Hodge-dual operator $*: \Omega^{k} (M)
\to \Omega^{6-k} (M)$, the chiral Lorentz generators $J_\pm^{AB} = \frac{1}{4}(\mathbf{I}_8 \pm
\Gamma_7)\Gamma^{AB} = \frac{1}{4}(\Gamma^{AB} \mp \frac{i}{4!} \varepsilon_{ABCDEF} \Gamma^{CDEF})$
correspond to the canonical split of the enlarged vector space \eq{doubling-6vec}.
Therefore, the two-forms $F^{(\pm)} = \frac{1}{2} F^{(\pm)}_{AB} e^{(\pm)A} \wedge e^{(\pm)B}$
in Eq. \eq{id-f} must be understood as the element of the irreducible vector space in Eq. \eq{split-6d}, i.e.,
\begin{equation}\label{ext-proj}
 F^{(\pm)} \in  \Omega_\pm^2 (M).
\end{equation}
As a result, $\mathcal{R}_{AB}$ on the left-hand side of Eq. \eq{id-f} has twice as many components as
the usual Riemann curvature tensor.

Let us summarize the gauge theory formulation in section 2.
Suppose that $J_*^{AB}$ and $T_*^a$ are Lie algebra generators in an irreducible representation $R_*$
of $Spin(6)$ and $SU(4)$, respectively. First consider an $SU (4)$ gauge field $B =B^a T_*^a = *C$
in the representation $R_*$ obtained by taking the Hodge dual of a four-form $C = C^a T_*^a$
and make the following identification:
\begin{eqnarray}\label{id2}
    && \widetilde{\omega} = \frac{1}{2} \widetilde{\omega}_{AB} J_*^{AB} \cong B = B^a T_*^a, \\
    \label{id3}
    && \theta = \frac{1}{2} \theta_{AB} J_*^{AB} \cong C = C^a T_*^a,
\end{eqnarray}
where $\widetilde{\omega} = \frac{1}{2} \widetilde{\omega}_{AB} J_*^{AB}$ is the dual spin
connection and $\widetilde{\omega}= * \theta$. Then the dual curvature tensors (\ref{dual-curvature})
and (\ref{4-form}) are, respectively, written as
\begin{eqnarray}\label{dual-gaugef1}
    && \widetilde{F} = dB + B \wedge B = * H, \\
    \label{dual-gaugef2}
    && H = (-)^{d-1} \delta C + C \barwedge C,
\end{eqnarray}
where $H$ is a four-form field strength whose Hodge dual is the field strength $\widetilde{F}$
in $SU(4)$ gauge theory.
The nilpotency of exterior differentials, $d^2 = \delta^2 = 0$,
immediately leads to the Bianchi identity
\begin{equation}\label{dual-bianchi}
    d \widetilde{F} + B \wedge \widetilde{F} - \widetilde{F} \wedge B = 0
    \qquad \Leftrightarrow  \qquad \delta H  + (-)^{d-1}
    \bigl( C \barwedge H - H \barwedge C \bigr) = 0.
\end{equation}
Hence the geometric structure described by the dual variables $({\widetilde{\omega}}_{AB},
\widetilde{R}_{AB})$ will be exactly parallel to the usual one described by $(\omega_{AB}, R_{AB})$.

Thus it is natural to put the two geometric structures on an equal footing. Moreover the irreducible
representation of the Clifford algebra $\mathbb{C}l (6)$ suggests that the curvature
tensors in Eq. (\ref{ext-proj}) are given by the combination
\begin{equation}\label{id-curvature}
    F^{(\pm)} = \frac{1}{2} (F \pm \widetilde{F}) = \frac{1}{2} (F \pm * H).
\end{equation}
One may note that, on an orientable (spin) manifold, the duplication of curvature tensors
always happens by the Hodge duality. The combination \eq{id-curvature} can be understood as follow.
One may regard the Riemann tensor $\mathcal{R}_{AB} = \frac{1}{2}\mathcal{R}_{ABCD} J^{CD}$
as a linear operator acting on the Hilbert space $V$ in Eq. \eq{spin-6}. As $\mathcal{R}_{AB}$ contains
two gamma matrices, it does not change the chirality of the vector space $V$.
Therefore, we can represent it in a subspace of definite chirality as
either $R^{(+)}_{AB}: S_+ \to S_+$ or $R^{(-)}_{AB}: S_- \to S_-$.
The former case $R^{(+)}_{AB}: S_+ \to S_+$ takes values in $\mathbf{4} \otimes
\overline{\mathbf{4}}$ in \eq{44*-clifford} with a singlet being removed
while the latter case $R^{(-)}_{AB}: S_- \to S_-$ takes values
in $\overline{\mathbf{4}} \otimes \mathbf{4}$ in \eq{4*4-clifford} with no singlet.
This implies two independent identifications defined by
\begin{eqnarray}\label{id-f+}
   && \mathbb{A}:  \frac{1}{2} R^{(+)}_{ABCD} J_+^{CD}
  \equiv F_{AB}^{(+)a}
   \big(T^a \oplus \mathbf{0}\big), \\
   \label{id-f-}
   && \mathbb{B}:  \frac{1}{2} R^{(-)}_{ABCD} J_-^{CD}
   \equiv F_{AB}^{(-)a}
   \big(\mathbf{0} \oplus (T^a)^*\big),
\end{eqnarray}
where the class $\mathbb{A} \; (\mathbb{B})$ acts on the subspace $S_+ \; (S_-)$
of positive (negative) chirality. See appendix A for the irreducible representation
of $Spin(6)$ and $SU(4)$. Because the classes $\mathbb{A}$
and $\mathbb{B}$ in Eqs. (\ref{id-f+}) and (\ref{id-f-}) are now represented by $4 \times 4$ matrices
on both sides, we can take a trace operation for the matrices which leads to the following relations
\begin{eqnarray}\label{id-fa}
   && \mathbb{A}:  R^{(+)}_{ABCD}  = - F_{AB}^{(+)a} \Tr (T^a J_+^{CD})
   \equiv F_{AB}^{(+)a} \eta^a_{CD}, \\
   \label{id-fb}
   && \mathbb{B}: R^{(-)}_{ABCD}  = - F_{AB}^{(-)a} \Tr \big( (T^{a})^* J_-^{CD} \big)
   \equiv F_{AB}^{(-)a} \overline{\eta}^{a}_{CD}.
\end{eqnarray}
Here we have introduced a six-dimensional analogue of the 't Hooft symbols defined by
\begin{equation}\label{6-thooft}
  \eta^{(\pm)a}_{AB} = - \Tr (T_\pm^a J_\pm^{AB}),
\end{equation}
where we used a bookkeeping notation, $\eta^{(+)a}_{AB} \equiv \eta_{AB}, \; \eta^{(-)a}_{AB}
\equiv \overline{\eta}_{AB}$ and $T_+^a \equiv T^a, \; T_-^a \equiv (T^a)^*$.
They serve as a complete basis of the vector space $\mathbf{15}$ in Eqs. \eq{44*-clifford} and \eq{4*4-clifford}.
An explicit expression of the six-dimensional 't Hooft symbols and
their algebra are presented in appendix B.

Note that $F^{(\pm)a} = \frac{1}{2} F_{AB}^{(\pm)a} e^{(\pm)A} \wedge e^{(\pm)B}$
in Eqs. \eq{id-f+} and \eq{id-f-} are the field strengths of $SU(4)$ gauge fields.
Thus we introduce a pair of $SU(4)$ gauge fields $\big(A^{(+)}, A^{(-)}\big)$
whose field strengths are given by
\begin{equation}\label{su4fs+-}
F^{(\pm)} = dA^{(\pm)} + A^{(\pm)} \wedge A^{(\pm)}.
\end{equation}
The $SU(4)$ gauge field $A^{(+)} \; (A^{(-)})$ is nothing but the spin connection
resident in the vector space $S_+ \; (S_-)$ of positive (negative) chirality, i.e.,
\begin{equation}\label{id-chiralspin}
  \omega^{(\pm)} = \frac{1}{2} \omega^{(\pm)}_{AB} J_\pm^{AB} \cong A^{(\pm)} = A^{(\pm)a} T_\pm^a.
\end{equation}
Using Eq. \eq{eta-1}, the field strengths can be written as $F_{AB}^{(\pm)a}
=  R^{(\pm)}_{ABCD} \eta^{(\pm)a}_{CD} = \eta^{(\pm)a}_{CD} R^{(\pm)}_{CDAB}$.
One can apply again the same expansion to the index pair $[AB]$ of the Riemann tensor $R^{(\pm)}_{CDAB}$.
That is, one can expand the $SU(4)$ field strengths in terms of the chiral bases in Eq. \eq{6-thooft}
\begin{eqnarray} \la{dec-fa}
\mathbb{A}:&&  F_{AB}^{(+)a} = f^{ab}_{(++)} \eta^b_{AB}, \\
\la{dec-fb}
\mathbb{B}:&& F_{AB}^{(-)a} = f^{ab}_{(--)} \overline{\eta}^{b}_{AB}.
\end{eqnarray}
As was pointed out in Eq. \eq{cliff-ext}, the Clifford algebra \eq{clifford-gen} is
isomorphic to the exterior algebra $\Lambda^* M$ as vector spaces, so the 't Hooft symbol
in Eq. \eq{6-thooft} has a one-to-one correspondence with the basis of two-forms
in $\Omega^2_\pm (M) = \Omega^2 (M) \oplus *\Omega^4 (M)$ depending on the chirality for a given orientation.
Consequently, the six-dimensional Riemann curvature tensors can be expanded as follows:
\begin{eqnarray} \la{class-i}
\mathbb{A}:&&  R^{(+)}_{ABCD} = f^{ab}_{(++)} \eta^a_{AB} \eta^b_{CD}, \\
\la{class-ii}
\mathbb{B}: && R^{(-)}_{ABCD} = f^{ab}_{(--)} \overline{\eta}^{a}_{AB} \overline{\eta}^{b}_{CD}.
\end{eqnarray}
Note that the index pairs $[AB]$ and $[CD]$ in the curvature tensor $R^{(\pm)}_{ABCD}$
have the same chirality structure because of the symmetry property $R^{(\pm)}_{ABCD} = R^{(\pm)}_{CDAB}$.

The Riemann curvature tensor in six dimensions has $225 = 15 \times 15$
components in total which is the number of the expansion coefficients $f^{ab}_{(\pm\pm)}$ in each class.
Because the torsion free condition has been assumed for the curvature tensors,
the first Bianchi identity $R^{(\pm)}_{A[BCD]}=0$ should be imposed which leads to 120 constraints
for each class. After all, the curvature tensor has $105 = 225-120$ independent
components which must be equal to the number of remaining expansion coefficients
in the class $\mathbb{A}$ or $\mathbb{B}$ after solving the 120 constraints
\begin{equation}\label{1st-bi}
 \varepsilon^{ABCEFG} R^{(\pm)}_{DEFG}=0.
\end{equation}
It is worthwhile to notice that the curvature tensor automatically satisfies
the symmetry property $R^{(\pm)}_{ABCD} = R^{(\pm)}_{CDAB}$ after dictating the first Bianchi
identity \eq{1st-bi}. Therefore, one can split the 120 constraints in Eq. \eq{1st-bi}
into the $105 = \frac{15 \times 14}{2}$ conditions imposing the symmetry
$R^{(\pm)}_{ABCD} = R^{(\pm)}_{CDAB}$ and the extra 15 conditions.
These extra conditions can be manifest by considering the tensor product of $SU(4)$ \ct{slansky}
\begin{equation}\label{tensor-prod}
    \mathbf{15} \otimes \mathbf{15} = (\mathbf{1} + \mathbf{15} + \mathbf{20}
    + \mathbf{84})_S \oplus (\mathbf{15} + \mathbf{45} + \overline{\mathbf{45}})_{AS}
\end{equation}
where the first part with 120 components is symmetric
and the second part with 105 components is antisymmetric. It is obvious from our
construction that $f^{ab}_{(\pm\pm)} \in \mathbf{15} \otimes \mathbf{15}$.
The 84 components in the symmetric part is the number of Weyl tensors in six dimensions
and the $21 = 20 + 1$ components refer to Ricci tensors.
The remaining 15 components in the symmetric part are removed by the first Bianchi identity \eq{1st-bi}
after expelling the antisymmetric components in Eq. \eq{tensor-prod}.

One can easily solve the symmetry property $R^{(\pm)}_{ABCD} = R^{(\pm)}_{CDAB}$ with the
coefficients satisfying
\begin{equation}\label{symm-coef}
 f^{ab}_{(++)} = f^{ba}_{(++)}, \qquad    f^{ab}_{(--)} =
 f^{ba}_{(--)},
\end{equation}
which results in 120 components for each chirality belonging to the symmetric part
in Eq. \eq{tensor-prod}. Now the remaining 15 conditions can be reduced to the equations
\begin{equation}\label{1st-bi-15}
 \varepsilon^{ABCDEF} R^{(\pm)}_{CDEF}=0.
\end{equation}
It is obvious that Eq. \eq{1st-bi-15} gives rise to a nontrivial relation
only for the coefficients satisfying Eq. \eq{symm-coef}.
Finally, using Eqs. \eq{eta-3} and \eq{eta-4}, Eq. \eq{1st-bi-15} can be reduced to the 15 constraints
\begin{equation}\label{1st-bianchi}
  d^{abc} f^{bc}_{(++)} = d^{abc} f^{bc}_{(--)} = 0
\end{equation}
for each sector. In the end, $f^{ab}_{(\pm\pm)}$ have 105 independent components for each chirality
which precisely match with the independent components
of Riemann curvature tensors in the class $\mathbb{A}$ or $\mathbb{B}$.\footnote{It may be worthwhile
to recall the four-dimensional situation \cite{ohya,loy}. In four dimensions, the first Bianchi identity
gives rise to 16 constraints. Thus Riemann curvature tensors have $20=36-16$ independent components.
And the 16 constraints split into 15 ones for $R_{ABCD} = R_{CDAB}$ and one more constraint which reads
as $\delta^{ab} f^{ab}_{(++)} = \delta^{\dot{a}\dot{b}} f^{\dot{a}\dot{b}}_{(--)}$.
The last constraint is responsible for the equality of the Ricci scalar $s$ in the chiral and
anti-chiral sectors in Eq. \eq{dec-riemann}. The constraints in Eq. \eq{1st-bianchi}
correspond to the six-dimensional analogue of the last one.}

Let us introduce the following (projection) operator acting on $6 \times 6$ antisymmetric
matrices defined by
\begin{equation} \label{so6-pro-op}
    P^{ABCD}_\pm \equiv \frac{1}{4} \big(\delta_{AC} \delta_{BD}
    - \delta_{AD} \delta_{BC} \big) \pm \frac{1}{8} \varepsilon^{ABCDEF} I_{EF}
    = P^{CDAB}_\pm
\end{equation}
where $I \equiv \mathbf{I}_3 \otimes i \sigma^2$. Because any $6 \times 6$
antisymmetric matrix of rank 4 spans a four-dimensional subspace $\mathbb{R}^4 \subset \mathbb{R}^6$,
the operator \eq{so6-pro-op} in this case can be written in the four-dimensional subspace as
\begin{equation} \label{so6-pro-op4}
    P^{ABCD}_\pm \equiv \frac{1}{4} \big(\delta_{AC} \delta_{BD}
    - \delta_{AD} \delta_{BC} \big) \pm \frac{1}{4} \varepsilon^{ABCD},
    \qquad (A,B,C,D) \in \mathbb{R}^4,
\end{equation}
so it reduces to the projection operator for such rank 4 matrices, i.e.,
\begin{equation}\label{4proj-id}
     P^{ABEF}_\pm  P^{EFCD}_\pm = P^{ABCD}_\pm,
     \qquad P^{ABEF}_\pm  P^{EFCD}_\mp = 0.
\end{equation}
Note that $I_{AB}$ is a $6 \times 6$ antisymmetric matrix of rank 6.
In this case, the operator \eq{so6-pro-op} does not act as a projection operator but acts as
\begin{equation} \label{so6-pro-op6}
    P^{ABCD}_\pm  I_{CD} = \Big( \frac{1}{2} \pm 1 \Big)
    I_{AB}.
\end{equation}
In general, one can deduce by a straightforward calculation the following properties
\begin{equation}\label{proj-genid}
     P^{ABEF}_\pm  P^{EFCD}_\pm = P^{ABCD}_\pm
     + \frac{1}{8} I_{AB} I_{CD},
     \qquad P^{ABEF}_\pm  P^{EFCD}_\mp = - \frac{1}{8} I_{AB} I_{CD}.
\end{equation}

After a little algebra, one can classify the 't Hooft symbols in Eq. \eq{6-thooft}
into the eigenspaces of the operator \eq{so6-pro-op}:
\begin{eqnarray} \la{su3+8}
 l^{(+)\hat{a}}_{AB} &\equiv& \left\{ \eta^{13}_{AB} = \frac{i}{2} \lambda_1 \otimes \sigma^2,
 \quad \eta^{14}_{AB} = \frac{i}{2} \lambda_2 \otimes \mathbf{I}_2, \quad \frac{1}{\sqrt{3}}
\Big(\eta^{8}_{AB} - \sqrt{2} \eta^{15}_{AB} \Big)
= -\frac{i}{2} \lambda_3 \otimes \sigma^2, \right. \xx
  && \quad \eta^{6}_{AB} = \frac{i}{2} \lambda_4 \otimes \sigma^2, \quad \eta^{7}_{AB}
  = -\frac{i}{2} \lambda_5 \otimes \mathbf{I}_2, \quad \eta^{11}_{AB}
  = \frac{i}{2} \lambda_6 \otimes \sigma^2, \\
  && \quad
  \left. \eta^{12}_{AB} = - \frac{i}{2} \lambda_7 \otimes \mathbf{I}_2, \quad \frac{2}{\sqrt{3}}
\Big(-\frac{1}{2} \eta^{3}_{AB} + \frac{1}{\sqrt{3}} \eta^8_{AB}
+ \frac{1}{\sqrt{6}} \eta^{15}_{AB} \Big) = - \frac{i}{2} \lambda_8 \otimes \sigma^2 \right\}, \nonumber \\
\la{+257}
m^{(+)\dot{a}}_{AB} &\equiv& \left\{ \eta^{1}_{AB} = \frac{i}{2} \lambda_2 \otimes \sigma^1, \quad
\eta^{2}_{AB} = -\frac{i}{2} \lambda_2 \otimes \sigma^3, \quad \eta^{9}_{AB}
= -\frac{i}{2} \lambda_5 \otimes \sigma^1, \right. \xx
  && \left. \quad \eta^{10}_{AB} = \frac{i}{2} \lambda_5 \otimes \sigma^3,
  \quad \eta^{4}_{AB} = \frac{i}{2} \lambda_7 \otimes \sigma^1, \quad
  \eta^{5}_{AB} = - \frac{i}{2} \lambda_7 \otimes \sigma^3 \right\},\\
\label{identity+1}
n^{(+)0}_{AB} &\equiv& \left\{ \eta^{3}_{AB} + \frac{1}{\sqrt{3}} \eta^8_{AB}
+ \frac{1}{\sqrt{6}} \eta^{15}_{AB}  = \frac{1}{2} I_{AB}
= \frac{1}{2} \mathbf{I}_3 \otimes  i \sigma^2 \right\},
\end{eqnarray}
where $\hat{a}, \hat{b} = 1, \cdots, 8$ and $\dot{a}, \dot{b} = 1, \cdots, 6$ are
$su(4)$ indices in the entries of $l^{(+)\hat{a}}_{AB}$ and $m^{(+)\dot{a}}_{AB}$,
respectively. They obey the following relations
\begin{eqnarray} \la{proj+lmn}
\begin{array}{ll}
 P_-^{ABCD} l^{(+)\hat{a}}_{CD} = l^{(+)\hat{a}}_{AB},
\qquad & P^{ABCD}_+ l^{(+)\hat{a}}_{CD} = 0, \\
P^{ABCD}_- m^{(+)\dot{a}}_{CD} =0,
\qquad  & P_+^{ABCD} m^{(+)\dot{a}}_{CD} = m^{(+)\dot{a}}_{AB}, \\
P_-^{ABCD} n^{(+)0}_{CD} = - \frac{1}{2} n^{(+)0}_{AB},
\qquad & P_+^{ABCD} n^{(+)0}_{CD} = \frac{3}{2} n^{(+)0}_{AB}.
\end{array}
\end{eqnarray}
Thus the (projection) operators \eq{so6-pro-op} decompose the vector space $\mathbf{15}$ into
their eigenspaces as $\mathbf{15}=\mathbf{8} \oplus \mathbf{6} \oplus \mathbf{1}$.

Similarly, one can also classify the 't Hooft symbols in Eq. \eq{6-thooft-b} into the eigenspaces
of the operator \eq{so6-pro-op}:
\begin{eqnarray} \la{su3-8}
 l^{(-)\hat{a}}_{AB} &\equiv& \left\{ - \overline{\eta}^{13}_{AB}
 = \frac{i}{2} \lambda_1 \otimes \sigma^2, \quad
\overline{\eta}^{14}_{AB} = \frac{i}{2} \lambda_2 \otimes \mathbf{I}_2, \quad \frac{1}{\sqrt{3}}
\Big(-\overline{\eta}^{8}_{AB} + \sqrt{2} \overline{\eta}^{15}_{AB} \Big)
= -\frac{i}{2} \lambda_3 \otimes \sigma^2, \right. \xx
  && \quad - \overline{\eta}^{9}_{AB} = \frac{i}{2} \lambda_4 \otimes \sigma^2, \quad
  -\overline{\eta}^{10}_{AB} = -\frac{i}{2} \lambda_5 \otimes \mathbf{I}_2, \quad
  \overline{\eta}^{4}_{AB} = \frac{i}{2} \lambda_6 \otimes \sigma^2, \xx
  && \quad
  \left. \overline{\eta}^{5}_{AB} = - \frac{i}{2} \lambda_7 \otimes \mathbf{I}_2,
  \quad \frac{2}{\sqrt{3}}
\Big(\frac{1}{2} \overline{\eta}^{3}_{AB} + \frac{1}{\sqrt{3}} \overline{\eta}^8_{AB}
+ \frac{1}{\sqrt{6}} \overline{\eta}^{15}_{AB} \Big) = - \frac{i}{2} \lambda_8 \otimes \sigma^2 \right\}, \\
\la{-257}
m^{(-)\dot{a}}_{AB} &\equiv& \left\{ - \overline{\eta}^{1}_{AB}
= \frac{i}{2} \lambda_2 \otimes \sigma^1, \quad \overline{\eta}^{2}_{AB}
= - \frac{i}{2} \lambda_2 \otimes \sigma^3, \quad - \overline{\eta}^{6}_{AB}
  = - \frac{i}{2} \lambda_5 \otimes \sigma^1,\right. \xx
  && \left. \quad - \overline{\eta}^{7}_{AB} = \frac{i}{2} \lambda_5 \otimes \sigma^3,
  \quad \overline{\eta}^{11}_{AB} = \frac{i}{2} \lambda_7 \otimes \sigma^1, \quad
   \overline{\eta}^{12}_{AB} = - \frac{i}{2} \lambda_7 \otimes \sigma^3 \right\},\\
\label{identity-1}
n^{(-)0}_{AB} &\equiv& \left\{ -\overline{\eta}^{3}_{AB} + \frac{1}{\sqrt{3}}
\overline{\eta}^8_{AB} + \frac{1}{\sqrt{6}} \overline{\eta}^{15}_{AB}
= \frac{1}{2} I_{AB} = \frac{1}{2} \mathbf{I}_3 \otimes i \sigma^2 \right\}.
\end{eqnarray}
The same properties such as Eq. \eq{proj+lmn} also hold for the above 't Hooft symbols.

The geometrical meaning of the (projection) operators in Eq. \eq{so6-pro-op} can be
understood as follows. Consider an arbitrary two-form\footnote{\label{supscrip+-}We will indicate
the superscript $(+)$ or $(-)$ only when we refer to a quantity belonging to a definite chirality class.
We will often omit the superscript whenever it is not necessary to specify the chirality class.}
\begin{equation}\label{two-form}
    F = \frac{1}{2} F_{MN} dx^M \wedge dx^N =  \frac{1}{2} F_{AB} e^A \wedge e^B
    \in \Omega^2 (M)
\end{equation}
and introduce the 15-dimensional complete basis of two-forms in $\Omega^2_\pm (M)$
for each chirality of $spin(6)$ Lorentz algebra
\begin{equation}\label{two-from}
    J_+^a \equiv \frac{1}{2} \eta^a_{AB} e^{(+)A} \wedge e^{(+)B} \in \Omega^2_+ (M), \qquad
    J_-^a \equiv \frac{1}{2} \overline{\eta}^a_{AB} e^{(-)A} \wedge e^{(-)A} \in \Omega^2_- (M).
\end{equation}
It is easy to derive the following identity using Eqs. \eq{eta-3} and \eq{eta-4}
\begin{equation}\label{vol-id}
    J^a_\pm \wedge J^b_\pm \wedge J^c_\pm = \frac{1}{2} d^{abc} \mathrm{vol} \big( g^{(\pm)} \big)
\end{equation}
where $\mathrm{vol} \big( g^{(\pm)} \big) = \sqrt{g^{(\pm)}} d^6 x$.
The Hodge-dual operator $*: \Omega^k (M) \to \Omega^{6-k} (M)$
is an isomorphism of vector spaces which depends upon a metric $g^{(\pm)}$
and the orientation of $M$. The nowhere vanishing volume form in (\ref{vol-id}) guarantees that
there exists a set of nondegenerate 2-forms on $M$
\begin{equation}\label{kahler-form}
\Omega_\pm = \frac{1}{2} I_{AB} e^{(\pm)A} \wedge e^{(\pm)B} = e^{(\pm)1} \wedge e^{(\pm)2}
+ e^{(\pm)3} \wedge e^{(\pm)4} + e^{(\pm)5} \wedge e^{(\pm)6}.
\end{equation}
This two-form can be wedged with the Hodge star to construct a diagonalizable
operator on $\Lambda^2 (M) = \Omega^2 (M) \oplus *\Omega^4 (M)$ as follows:
\begin{equation}\label{w-hodge}
    *_{\Omega_\pm} \equiv *(\bullet \wedge \Omega_\pm): \Omega^2 (M)
    \stackrel{\bullet \wedge \Omega_\pm}{\longrightarrow} \Omega^4 (M)
    \stackrel{*}{\longrightarrow} \Omega^2 (M)
\end{equation}
by $*_{\Omega_\pm} (\alpha) = *(\alpha \wedge \Omega_\pm)$ for $\alpha \in \Omega^2 (M)$.
After a little inspection, the $15 \times 15$ matrix representing $*_{\Omega_\pm}$ is found to have
the eigenvalues $2, 1$ and $-1$ with the eigenspaces of dimension 1, 6 and 8, respectively.
On any six-dimensional orientable spin manifold $M$, the space of 2-forms
$\Omega^2_+ (M)$ in the positive chirality space can thus be decomposed
into three subspaces
\begin{equation}\label{2-from-dec}
    \Omega^2_+ (M) = \Lambda^2_1 \oplus \Lambda^2_6 \oplus \Lambda^2_8,
\end{equation}
which coincides with the decomposition in Eq. \eq{proj+lmn}.
The spaces $\Lambda^2_1$ and $\Lambda^2_6$ are locally spanned by
\begin{eqnarray} \label{basis-1}
 \Lambda^2_1 &=& \Omega_+, \\
 \label{basis-6}
    \Lambda^2_6 &=& \left\{ J_+^1, J_+^2, J_+^4, J_+^5, J_+^9, J_+^{10} \right\},
\end{eqnarray}
and $\Lambda^2_8$ by
\begin{eqnarray}  \label{basis-8}
    \Lambda^2_8 &=& \left\{ J_+^6, J_+^7, J_+^{11}, J_+^{12}, J_+^{13}, J_+^{14},
   K_+, L_+  \right\}
\end{eqnarray}
with $K_+ \equiv \frac{1}{\sqrt{3}} \big(J_+^{8} - \sqrt{2} J_+^{15} \big)$ and
$L_+ \equiv \frac{2}{\sqrt{3}} \big(-\frac{1}{2} J_+^{3} + \frac{1}{\sqrt{3}} J_+^8
+ \frac{1}{\sqrt{6}} J_+^{15} \big)$. A similar decomposition can be done with
the negative chirality basis $J_-^a$.

Note that the entries of $\Lambda^2_1,  \Lambda^2_6$ and $\Lambda^2_8$ coincide
with those of $n^{(\pm)0}_{AB}, m^{(\pm)\dot{a}}_{AB}$ and $l^{(\pm)\hat{a}}_{AB}$,
respectively. One can quickly see that this coincidence is not an accident.
Consider the action of the projection operator \eq{so6-pro-op} on
the two-form \eq{two-form}, which is given by
\begin{equation}\label{proj-act}
    P^{ABCD}_\pm F_{CD} = \frac{1}{2} \big(F_{AB} \pm \frac{1}{4} \varepsilon^{ABCDEF}
    F_{CD} I_{EF} \big)
\end{equation}
or in terms of form notation
\begin{equation}\label{act-proj-form}
2 P_\pm F = F \pm * (F \wedge \Omega_\pm) = F \pm *_{\Omega_\pm} F.
\end{equation}
It is easy to see that $F \in \Lambda^2_8$ if $P_+ F=0$, so it satisfies
the $\Omega$-anti-self-duality equation
\begin{equation}\label{omega-asd}
   * (F \wedge \Omega_\pm) = - F,
\end{equation}
whereas $F \in \Lambda^2_6$ satisfies the $\Omega$-self-duality equation $P_- F=0$, i.e.,
\begin{equation}\label{omega+asd}
   * (F \wedge \Omega_\pm) = F.
\end{equation}

It is not difficult to show \ct{string-book} that the set $\big\{l^{(\pm)\hat{a}}_{AB}, n^{(\pm)0}_{AB} \big\}$
can be identified with $u(3)$ generators which are embedded in $so(6) \cong su(4)$.
In general, an element of $U(3)$ group can be represented as
\begin{equation}\label{u3}
U =  \exp\Big(i \sum_{a=0}^8 \theta^{a} \lambda_{a}\Big) \equiv e^\Theta
\end{equation}
where $\lambda_0 = \mathbf{I}_3$ is a $3 \times 3$ unit matrix, $\lambda_{\hat{a}} \;
(\hat{a} = 1, \cdots, 8)$ are the $su(3)$ Gell-Mann matrices and $\theta^{a}$'s
are real parameters for $U$ to be unitary. The $3 \times 3$ anti-Hermitian matrix
$\Theta$ consists of matrix elements which are complex numbers $\Theta_{i\overline{j}} =
- (\Theta_{\overline{j}i})^* \; (i,\overline{j}=1,2,3)$ and it can easily be embedded into a $6 \times 6$
real matrix in $so(6)$ Lie algebra by replacing $\Theta_{i\overline{j}} = \mathrm{Re} \Theta_{i\overline{j}} + i \mathrm{Im} \Theta_{i\overline{j}}$ by the $2 \times 2$ real matrix $\widetilde{\Theta}_{AB} = \mathbf{I}_2 \cdot \mathrm{Re} \Theta_{i\overline{j}}
+ i  \sigma^2 \cdot \mathrm{Im} \Theta_{i\overline{j}}$.
A straightforward calculation (see Eq. (\ref{complex-thooft})) shows that the resulting $6 \times 6$
antisymmetric real matrix $\widetilde{\Theta}_{AB}$ can be written as
\begin{equation}\label{so6-u3}
\widetilde{\Theta}_{AB} = 2 \big( \theta^0 n^{(\pm)0}_{AB} +
\theta^{\hat{a}} l^{(\pm)\hat{a}}_{AB} \big) =
P^{ABCD}_- \widetilde{\Theta}_{CD} + 3 \theta^0 n^{(\pm)0}_{AB}.
\end{equation}

Note that $U(3)$ is the holonomy group of K\"ahler manifolds.
That is, the projection operators in Eq. \eq{so6-pro-op} can serve to project a Riemannian
manifold whose holonomy group is $SO(6)$ into a K\"ahler manifold with $U(3)$ holonomy.
Let us show that it is indeed the case. Suppose that $M$ is a complex manifold.
Let us introduce local complex coordinates $z^\alpha =\{x^1 + i x^2, x^3 + i x^4, x^5 + i x^6 \},
\; \alpha=1,2,3$ and their complex conjugates $\bar{z}^{\bar{\alpha}}, \; \bar{\alpha}=1,2,3$,
in which a complex structure $J$ takes the form ${J^\alpha}_\beta
= i {\delta^\alpha}_\beta, \; {J^{\bar{\alpha}}}_{\bar{\beta}}
= - i {\delta^{\bar{\alpha}}}_{\bar{\beta}}$ \ct{string-book}.
Note that, relative to the real basis $x^M, M=1, \cdots, 6$, the complex structure
is given by $J= I = \mathbf{I}_3 \otimes  i \sigma^2$ which was already
introduced in Eq. \eq{so6-pro-op}. We further impose the Hermitian condition
on the complex manifold $M$ defined by $g(X,Y) = g (JX,JY)$ for any $X,Y \in \Gamma(TM)$.
This means that the Riemannian metric $g$ on the complex manifold $M$ is a Hermitian metric,
i.e. $g_{\alpha\beta} = g_{\bar{\alpha}\bar{\beta}} = 0, \; g_{\alpha\bar{\beta}}
= g_{\bar{\beta}\alpha}$. The Hermitian condition can be solved by taking the vielbeins as
\begin{equation}\label{holo-vielbein}
    e^i_{\bar{\alpha}} = e^{\bar{i}}_{\alpha} = 0 \quad {\rm and} \quad
    E_i^{\bar{\alpha}} = E_{\bar{i}}^{\alpha} = 0
\end{equation}
where a tangent space index $A=1,\cdots,6$ has been split into a holomorphic index $i=1,2,3$
and an anti-holomorphic index $\bar{i}=1,2,3$. This in turn means that
${J^i}_j = i {\delta^i}_j, \; {J^{\bar{i}}}_{\bar{j}} = - i {\delta^{\bar{i}}}_{\bar{j}}$.
Then one can see that the two-form $\Omega_\pm$ in Eq. \eq{kahler-form} is a K\"ahler form,
i.e., $\Omega_\pm(X,Y) = g^{(\pm)}(JX,Y)$ and it is given by
\begin{equation}\label{kahler}
    \Omega_\pm = i e^{(\pm)i} \wedge e^{(\pm)\bar{i}} = i e^{(\pm)i}_{\alpha}
    e^{(\pm)\bar{i}}_{\bar{\beta}}
    dz^\alpha \wedge d\bar{z}^{\bar{\beta}} =
    i g^{(\pm)}_{\alpha \bar{\beta}} dz^\alpha \wedge d\bar{z}^{\bar{\beta}}
\end{equation}
where $e^{(\pm)i} = e^{(\pm)i}_\alpha dz^\alpha$ is a holomorphic one-form and
$e^{(\pm)\bar{i}} = e^{(\pm)\bar{i}}_{\bar{\alpha}} d\bar{z}^{\bar{\alpha}}$ is
an anti-holomorphic one-form. It is also easy to see that the condition for a Hermitian
manifold $(M, g^{(\pm)})$ to be K\"ahler, i.e. $d \Omega_\pm = 0$,
is equivalent to the one that the spin connection $\omega^{(\pm)}_{AB}$ is $U(3)$-valued, i.e.,
\begin{equation}\label{spin-u3}
    \omega^{(\pm)}_{ij} = \omega^{(\pm)}_{\bar{i}\bar{j}} = 0.
\end{equation}
Therefore, the spin connection after the K\"ahler condition \eq{spin-u3} can be written
as the form \eq{so6-u3}.

All the above results can be clearly understood by the properties of $Spin(6)$ and $SU(4)$ groups.
Introducing complex coordinates on $\mathbb{R}^6$ means that one has to consider the Lorentz
subgroup $U(3) \subset SU(4)$ acting on $\mathbb{C}^3 \subset \mathbb{C}^4$ and
so one decomposes the $\mathbf{4}$ and $\overline{\mathbf{4}}$ of $SU(4)$
as $\mathbf{4} = \mathbf{1}_1 \oplus \mathbf{3}_{-\frac{1}{3}}$ and $\overline{\mathbf{4}}
= \overline{\mathbf{1}}_{-1} \oplus \overline{\mathbf{3}}_{\frac{1}{3}}$
under $U(3)=U(1) \times SU(3)$ where the subscripts denote $U(1)$ charges.
Using the branching rule of $SU(4) \supset U(1) \times SU(3)$ \ct{slansky}, one can get
the following decompositions after removing $SU(4)$ singlets
\begin{eqnarray} \la{branching1}
&& \mathbf{4} \otimes \overline{\mathbf{4}} - \mathbf{1}
= (\mathbf{3} \otimes \overline{\mathbf{3}})_0
\oplus (\mathbf{3}_{-\frac{4}{3}} \oplus \overline{\mathbf{3}}_{\frac{4}{3}} )
= (\mathbf{8} \oplus \mathbf{1})_0
\oplus (\mathbf{3}_{-\frac{4}{3}} \oplus \overline{\mathbf{3}}_{\frac{4}{3}} ),  \\
\la{branching2}
&& \overline{\mathbf{4}} \otimes \mathbf{4} - \mathbf{1}
= (\overline{\mathbf{3}} \otimes \mathbf{3})_0
\oplus (\mathbf{3}_{-\frac{4}{3}} \oplus \overline{\mathbf{3}}_{\frac{4}{3}} )
= (\mathbf{8} \oplus \mathbf{1})_0
\oplus (\mathbf{3}_{-\frac{4}{3}} \oplus \overline{\mathbf{3}}_{\frac{4}{3}} ).
\end{eqnarray}
The spin connection $\omega_{AB} \in \mathbf{15}$ can be decomposed
according to the above branching rule as
\begin{eqnarray} \la{spin-branching}
\begin{array}{ll}
\omega_{ij} \in \mathbf{3}_{-\frac{4}{3}}, \qquad &\omega_{\bar{i}\bar{j}}
\in \overline{\mathbf{3}}_{\frac{4}{3}}, \\
\omega_{\bar{i}j} - \frac{1}{3} \delta^i_j \omega_{\bar{k}k} \in \mathbf{8}_{0},
\qquad &\omega_{\bar{i}i} \in \mathbf{1}_{0}.
\end{array}
\end{eqnarray}
Hence the K\"ahler condition \eq{spin-u3} means that spin connections in
$\mathbf{3}_{-\frac{4}{3}}$ and $\overline{\mathbf{3}}_{\frac{4}{3}}$ decouple
from the theory and only the components in $\mathbf{8}_{0}$ and $\mathbf{1}_0$ survive.
It is now obvious why we could have such decompositions in Eqs. \eq{su3+8}-\eq{identity-1}
in which $l^{(\pm)\hat{a}}_{AB} \in \mathbf{8}_0, \; m^{(\pm)\dot{a}}_{AB}
\in (\mathbf{3}_{-\frac{4}{3}} \oplus \overline{\mathbf{3}}_{\frac{4}{3}} )$
and $n^{(\pm)0}_{AB} \in \mathbf{1}_0$.

One can rephrase the K\"ahler condition \eq{spin-u3} using the gauge theory formalism.
From the identification $\omega^{(\pm)} \equiv \Gamma_\pm \varpi = A^{(\pm)a} T^a_\pm$
in Eq. \eq{id-chiralspin}, we get the relation
\begin{equation} \label{id-gauge-spin}
    \omega^{(\pm)}_{AB} = A^{(\pm)a} \eta^{(\pm)a}_{AB}, \qquad
    A^{(\pm)a} = - 2 \mathrm{Tr} \big(\omega^{(\pm)} T^a_\pm \big).
\end{equation}
We will focus on the type $\mathbb{A}$ case as the same analysis can be applied
to the type $\mathbb{B}$ case. If $(M, g^{(+)})$ is a K\"ahler manifold, Eq. \eq{spin-u3} means that
\begin{equation}\label{0A}
  A^{(+)1} = A^{(+)2} = A^{(+)4} = A^{(+)5} = A^{(+)9} = A^{(+)10} = 0
\end{equation}
because $\eta^{a}_{ij} \neq 0$ only for $a= 1,2,4,5,9,10$, otherwise $\eta^{a}_{ij} = 0$.
See Eq. \eq{complex-thooft}. This result is consistent with the branching rule \eq{spin-branching},
i.e., $m^{(+)\dot{a}}_{AB} \in (\mathbf{3}_{-\frac{4}{3}} \oplus \overline{\mathbf{3}}_{\frac{4}{3}} )$.
In other words,  $A^{(+)\dot{a}} = 0$ and so the gauge fields take values in $u(3)$ Lie algebra
according to the result \eq{so6-u3}.
Then the $SU(4)$ structure constants $f^{abc}$ in the Table \ref{table-sf} guarantee
that the corresponding field strengths also vanish, i.e.,
\begin{equation} \label{vanishing-f}
F^{(+)a} = \frac{1}{2} f^{ab}_{(++)} \eta^b_{AB} e^{(+)A} \wedge e^{(+)B}= 0
\end{equation}
for $a= 1,2,4,5,9,10$. Thus we get $f^{\dot{a}b}_{(++)} = 0$ for $\forall \, b=1, \cdots, 15$.
This immediately leads to the conclusion that
\begin{eqnarray}\label{hym}
  F^{(+)a} &=& d A^{(+)a} - \frac{1}{2} f^{abc} A^{(+)b} \wedge A^{(+)c} \xx
  &=& f^{ab}_{(++)} J_+^b
  \in \Lambda^2_8 \oplus \Lambda^2_1
\end{eqnarray}
where $F^{(+)a}, \; a = 0, 1, \cdots, 8$, are the field strengths of $U(3)$ gauge fields.
As will be shown below, $F^{(+)0} \in \Lambda^2_1$ is the field strength of
the $U(1)$ part of $U(3)$ spin connections
and $F^{(+)\hat{a}} \in \Lambda^2_8, \; \hat{a} = 1, \cdots, 8$, belong to the $SU(3)$ part.
In particular, as $F^{(+)\hat{a}} \in \Lambda^2_8$, they satisfy
the $\Omega$-anti-self-duality equation \eq{omega-asd} known as
the HYM equation \ct{duy}
\begin{equation}\label{duy}
    F^{(+)\hat{a}} = - * (F^{(+)\hat{a}} \wedge \Omega_+), \qquad
    \hat{a} = 1, \cdots, 8.
\end{equation}

It is well-known \ct{string-book} that the Ricci-tensor of a K\"ahler manifold is
the field strength of the $U(1)$ part of the $U(3)$ spin connection.
Therefore, the Ricci-flat condition can be stated as $F^{(+)0} = 0$. One can explicitly check it as follows.
Recall that $F^{(+)a}_{AB} = f^{ab}_{(++)} \eta^b_{AB}$. Using the result \eq{vanishing-f},
one can see that the nonzero components of $f^{ab}_{(++)}$ run only over $(a,b) \in
\{3,6,7,8,11,12,13,14,15\}$. Thereby the constraint \eq{1st-bianchi}  becomes nontrivial
only for those values. As a result, the number of independent components of
$f^{ab}_{(++)}$ is given by $\frac{9 \times 10}{2} - 9 = 36$. The Ricci-flat
condition $R^{(+)}_{AB} \equiv R^{(+)}_{ACBC} = f^{ab}_{(++)} \eta^a_{AC} \eta^b_{BC} = 0$
further constrains the coefficients. A close inspection shows that out of 21 equations, $R^{(+)}_{AB}=0$,
only 9 equations are independent and, after utilizing the constraint \eq{1st-bianchi},
the equations for the Ricci-flatness can be succinctly arranged as
\begin{equation}\label{u1=0}
    f_{(++)}^{3a} + \frac{1}{\sqrt{3}} f_{(++)}^{8a} + \frac{1}{\sqrt{6}} f_{(++)}^{15a} = 0.
\end{equation}
The above condition means that
\begin{eqnarray}\label{u1c=0}
    F^{(+)0}_{AB} &=& \Big(f_{(++)}^{3a} + \frac{1}{\sqrt{3}} f_{(++)}^{8a}
    + \frac{1}{\sqrt{6}} f_{(++)}^{15a} \Big)
    \eta^a_{AB}  \xx
    &=&  F^{(+)3}_{AB} + \frac{1}{\sqrt{3}}  F^{(+)8}_{AB}
    + \frac{1}{\sqrt{6}}  F^{(+)15}_{AB} = 0.
\end{eqnarray}
If one introduces a gauge field defined by
\begin{equation}\label{u1-gauge}
    A^{(+)0} \equiv \omega^{(+)}_{AB} n^{(+)0}_{AB} = A^{(+)3} + \frac{1}{\sqrt{3}}  A^{(+)8}
    + \frac{1}{\sqrt{6}}  A^{(+)15},
\end{equation}
one can show that the field strength in Eq. \eq{u1c=0} is given by
\begin{equation}\label{u1-curvature}
  F^{(+)0} = d A^{(+)0}
\end{equation}
after using the fact that the $U(3)$ structure constants $f^{abc}$
satisfy the following relation
\begin{equation}\label{u3+structure}
    f^{3ab} + \frac{1}{\sqrt{3}} f^{8ab} + \frac{1}{\sqrt{6}} f^{15ab} = 0.
\end{equation}
The relation \eq{u3+structure} is easy to understand because the $U(1)$ part among
the $U(3)$ structure constants has to vanish.
This establishes the result that the Ricci-flatness is equal to the vanishing of
the $U(1)$ field strength. That is, $F^{(+)0} = d A^{(+)0} \in \Lambda^2_1$ has
a trivial first Chern class.

The same result can be obtained for the type $\mathbb{B}$ case.
The K\"ahler condition \eq{spin-u3} can similarly be solved by
\begin{equation}\label{0A-}
  A^{(-)1} = A^{(-)2} = A^{(-)6} = A^{(-)7} = A^{(-)11} = A^{(-)12} = 0.
\end{equation}
Note that the entries of $U(3)$ generators for the type $\mathbb{B}$ case are different from
those for the type $\mathbb{A}$ case.
The Ricci-flat condition $R^{(-)}_{AB} \equiv R^{(-)}_{ACBC} = f^{ab}_{(--)} \overline{\eta}^a_{AC}
\overline{\eta}^b_{BC} = 0$ leads to the equation
\begin{equation} \label{u1-=0}
    - f_{(--)}^{3a} + \frac{1}{\sqrt{3}} f_{(--)}^{8a} + \frac{1}{\sqrt{6}}
    f_{(--)}^{15a} = 0.
\end{equation}
It is equivalent to the vanishing of $U(1)$ field strength, i.e. $F^{(-)0} = d A^{(-)0} = 0$,
where the $U(1)$ gauge field is defined by
\begin{equation}\label{-u1-gauge}
    A^{(-)0} \equiv \omega^{(-)}_{AB} n^{(-)0}_{AB} = - A^{(-)3} + \frac{1}{\sqrt{3}}  A^{(-)8}
    + \frac{1}{\sqrt{6}}  A^{(-)15}.
\end{equation}
This fact can be derived by using the fact that the $U(3)$ structure constants $f^{abc}$
for the type $\mathbb{B}$ case satisfy the following relation
\begin{equation}\label{u3-structure}
   - f^{3ab} + \frac{1}{\sqrt{3}} f^{8ab} + \frac{1}{\sqrt{6}} f^{15ab} = 0
\end{equation}
where $a,b$ now run over $3,4,5,8,9,10,13,14,15$.
Hence one can see that CY manifolds for the type $\mathbb{B}$ case also obey the HYM equations
\begin{equation}\label{duy-}
    F^{(-)\hat{a}} = - * (F^{(-)\hat{a}} \wedge \Omega_-), \qquad
    \hat{a} = 1, \cdots, 8.
\end{equation}

Note that the HYM equations for a vector bundle $E$ over a CY manifold $M$ define a set of
differential equations satisfied by the gauge fields of the vector bundle $E \to M$.
In general, the connection of a vector bundle $E \to M$ over a CY manifold $M$
is not related to the spin connection of the CY manifold $M$ unless $E = TM$.
For example, the $G$-bundle over a CY manifold $M$ is such a case.
However, in our case, we have adopted the so-called standard
embedding $E = TM$ where $TM$ is the tangent bundle of a CY manifold $M$.
Then the gauge fields of $E=TM$ are the spin connections on the tangent bundle $TM$ of the CY manifold $M$.
Since the tangent bundle $TM$ is defined by the CY manifold $M$ itself,
the HYM instanton in this case is inherited from the CY manifold.
Therefore the HYM instanton for the tangent bundle $TM$ cannot be identified with an ordinary
Yang-Mills instanton on a {\it fixed} background manifold since the Yang-Mills connection of $TM$
is directly determined by the CY manifold $M$.

In summary, the K\"ahler condition \eq{spin-u3} projects the 't Hooft symbols
to $U(3)$-valued ones in $\mathbf{1}_0 \oplus \mathbf{8}_0$
and results in the reduction of the gauge group
from $SU(4)$ to $U(3)$. The Ricci-flatness is equivalent to the condition
$F^{(\pm)0} = d A^{(\pm)0} = 0 \in \mathbf{1}_0$, so the gauge group is further reduced to $SU(3)$.
Remaining spin connections are $SU(3)$ gauge fields that belong to $\mathbf{8}_0$ and satisfy
the HYM equations \eq{duy} or \eq{duy-}. As a K\"ahler manifold with the trivial
first Chern class is a CY manifold, we see that the CY condition is equivalent
to the HYM equations whose solution is known as HYM instantons \ct{string-book}.
Consequently, we find that a six-dimensional CY manifold automatically satisfies the
HYM equations in $SU(3)$ Yang-Mills gauge theory, but the converse is not generally true.

\section{Mirror Symmetry of Calabi-Yau Manifolds}

In this section we want to explore the geometrical properties of six-dimensional
Riemannian manifolds in the irreducible representations $\mathbb{A}$ and $\mathbb{B}$.
In section 2, we have introduced dual vielbeins $\widetilde{e}^A = (*h)^A$ and dual spin
connections $\widetilde{\omega}_{AB} = (* \theta)_{AB}$ in addition to usual
ones $(e^A, \omega_{AB})$. The dual geometric structure described
by $(\widetilde{e}^A, \widetilde{\omega}_{AB}) \cong \big( h^A, \theta_{AB} \big)$ is basically
originated from the Hodge duality of the exterior algebra $\Lambda^* M$ on an orientable manifold $M$.
According to the chiral structure of irreducible representations
in Eqs. (\ref{44*-clifford})-(\ref{4*4*-clifford}), we have associated two geometric structures
$\big(e^{(+)A}, \omega^{(+)}_{AB}\big)$ and $\big(e^{(-)A}, \omega^{(-)}_{AB}\big)$
on a spin manifold $M$ where
\begin{eqnarray} \label{ab-viel}
 && e^{(\pm)A} = \frac{1}{2} (e \pm \widetilde{e})^A = \frac{1}{2} (e \pm * h)^A, \\
 \label{ab-spinc}
 && \omega^{(\pm)}_{AB} = \frac{1}{2} (\omega \pm \widetilde{\omega})_{AB}
 = \frac{1}{2} (\omega \pm * \theta)_{AB}.
\end{eqnarray}
This means that there are two independent ways to characterize a six-dimensional spin manifold.
Accordingly we can consider two kinds of Riemannian manifolds depicted by the metrics
\begin{equation}\label{two-metric}
    ds_{\mathbb{A}}^2 = e^{(+)A} \otimes e^{(+)A}, \qquad ds_{\mathbb{B}}^2
    = e^{(-)A} \otimes ^{(-)A},
\end{equation}
where $\mathbb{A}$ and $\mathbb{B}$ refer to their chirality class.
Each of the metrics defines their own spin connections $\omega^{(\pm)}_{AB} = \omega^{(\pm)}_{AB}
\big( e^{(\pm)} \big)$ through the torsion-free condition \eq{double-torsion}.
Generally speaking, the six-dimensional spin manifolds described by the $\mathbb{A}$ and $\mathbb{B}$
metrics \eq{two-metric} are indepedent of each other, so the variety is simply doubled
due to the Hodge duality on $\Lambda^* M$.

The spin connections can take arbitrary values as far as they satisfy the integrability
condition \eq{2-1bianchi}. Their symmetry properties can be characterized by decomposing them
into the irreducible subspaces under $SO(6)$ group:
\begin{equation} \la{spin-so6-young}
\Yvcentermath1
{\tiny \omega_{ABC} \in \mathbf{6} \otimes \mathbf{15} = \yng(1) \otimes \yng(1,1)
= \yng(1,1,1)  \oplus  \yng(2,1)} = \mathbf{20} \oplus \mathbf{70}
\end{equation}
where ${\tiny  \Yvcentermath1 \yng(1,1,1)} = \mathbf{20}$ is a completely antisymmetric part
of spin connections defined by $\omega_{[ABC]} =
\frac{1}{3}(\omega_{ABC} + \omega_{BCA} + \omega_{CAB})$.
In six dimensions, the spin connections $\omega_{[ABC]}$ may be further decomposed
into (imaginary) self-dual (sd) and anti-self-dual (asd) parts, i.e.,
\begin{equation}\label{sd-3-form}
\omega_{[ABC]} = \left(\omega^{sd}_{[ABC]} \in \mathbf{10} \right) \oplus
\left(\omega^{asd}_{[ABC]} \in \mathbf{10} \right).
\end{equation}
The above decomposition may be shaky because ${\tiny  \Yvcentermath1 \yng(1,1,1)} = \mathbf{20}$
is already an irreducible representation of $SO(6)$.
It is just for a heuristic comparison with the irreducible $SU(4)$ representation.
Note that $\mathbf{6}$ is coming from the antisymmetric tensor in
$\mathbf{4} \times \mathbf{4}$ in Eq. \eq{44-clifford} or
$\overline{\mathbf{4}} \times \overline{\mathbf{4}}$ in Eq. \eq{4*4*-clifford}.
Thus, under $SU(4)$ group,
one can instead get the following decomposition of the spin connections \ct{slansky}
\begin{equation} \la{spin-so6-dec}
\omega_{ABC} \in \mathbf{6} \otimes \mathbf{15} = {\tiny  \Yvcentermath1 \yng(1,1)} \otimes
{\tiny  \Yvcentermath1 \yng(2,1,1)} =  \left(\mathbf{6}
= {\tiny  \Yvcentermath1 \yng(2,2,1,1)} \right) \oplus \left(\mathbf{10}
= {\tiny  \Yvcentermath1 \yng(3,1,1,1)} \right)  \oplus \left(\overline{\mathbf{10}}
= {\tiny  \Yvcentermath1 \yng(2,2,2)} \right)  \oplus \left( \mathbf{64}
= {\tiny  \Yvcentermath1 \yng(3,2,1)} \right).
\end{equation}
Hence notice that the irreducible representation of $SU(4)$ for spin connections is
more refined than the irreducible {\it spinor} representation of $SO(6)$.

Given a metric $ds^2 = e^A \otimes e^A$, one can determine the spin connection $\omega_{AB}$
using the torsion free condition, $T^A = de^A + {\omega^A}_B \wedge e^B =0$.
Because we are dictating an irreducible spinor representation of local Lorentz symmetry
for the identification \eq{id}, it is necessary to specify which representation
is chosen to embed the spin connection $\omega_{AB}$.
One can equally choose either the positive or negative chiral representation.
This situation may be familiar with a supersymmetric solution in supergravity.
To be specific, consider the supersymmetry transformation of six-dimensional gravitino $\Psi_M$
given by $\delta \Psi_M = D_M \eta$ where a Dirac operator $D_M = \partial_M + \omega_M$
acts on a chiral spinor $\eta$. Then a background geometry obeying
$\delta \Psi_M = D_M \eta = 0$ must satisfy a well-known condition $[D_M, D_N] \eta
= \frac{1}{2} R_{MNPQ} J^{PQ} \eta = 0$. In this case the representation is determined by an unbroken
supersymmetry generated by the chiral spinor $\eta$.
Hence the corresponding $SU(4)$ gauge fields are also identified according to
the map \eq{id-chiralspin}, depending on the chiral representation chosen by
the supersymmetric background geometry.
Whenever a metric is known in a specific chiral representation,
one can determine the coefficients $f^{ab}_{(++)}$ in Eq. \eq{class-i} or $f^{ab}_{(--)}$
in Eq. \eq{class-ii} through the explicit calculation of Riemann curvature tensors.
Since the geometric structures described by the data
$\big(e^{(+)A}, \omega^{(+)}_{AB} \big)$ and $\big(e^{(-)A}, \omega^{(-)}_{AB}\big)$ are
completely independent of each other, one can attribute them to two different Riemannian manifolds.

So let us denote the geometric structures $\big(e^{(+)A}, \omega^{(+)}_{AB} \big)$ and $\big(e^{(-)A}, \omega^{(-)}_{AB}\big)$ by $\mathbb{A}$ and $\mathbb{B}$, respectively, according to Eq. \eq{pair-data}.
Suppose that the geometric structures $(\mathbb{A}, \mathbb{B})$ describe a pair of six-dimensional
spin manifolds $(M_+, M_-)$. Since each manifold can be described by either $\mathbb{A}$-type
or $\mathbb{B}$-type,
the geometric data for the pair are given by either $(M_+ (\mathbb{A}), M_-(\mathbb{B}))$ or
$(M_+ (\mathbb{B}), M_-(\mathbb{A}))$.
This pairing has shown up in the Table \ref{table-ms}.
In general, the manifolds $(M_+, M_-)$ in the pair are assumed to be different
even topologically. Given their metrics for the pair, one can determine the coefficients
$\big( f^{ab}_{(++)}, f^{ab}_{(--)} \big)$ in Eqs. (\ref{class-i}) and (\ref{class-ii}).
Since the pair consist of independent manifolds, it is possible
to arrange the pair such that the coefficients $\big( f^{ab}_{(++)}, f^{ab}_{(--)} \big)$
obey some relation, e.g. Eq. (\ref{mirror-cyh1}). To be specific, let us choose the embedding
$(M_+ (\mathbb{A}), M_-(\mathbb{B}))$.
Thus one CY manifold $M_+ = M$ is described by the metric $ds_{\mathbb{A}}^2 = e^{(+)A} \otimes e^{(+)A}$
of the type $\mathbb{A}$ while the other CY manifold $M_- = \widetilde{M}$ is described by the metric
$ds_{\mathbb{B}}^2 = e^{(-)A} \otimes e^{(-)A}$ of the type $\mathbb{B}$.
The Euler characteristic $\chi(M)$ of a six-dimensional Riemannian manifold $M$ is defined by
\begin{equation}\label{euler-betti}
    \chi(M) = \sum_{r=0}^6 (-)^r b_r
\end{equation}
where $b_r = \sum_{p+q=r} h^{p,q} (M)$ is the $r$-th Betti number.
A mirror pair of CY manifolds $(M, \widetilde{M})$ obeys the property $h^{p,q} (M) =
h^{3-p,q}(\widetilde{M})$. This property leads to an important result that the Euler characteristic
of the mirror manifold $\widetilde{M}$ has an opposite sign, i.e., $\chi(\widetilde{M}) = - \chi (M)$.
Thus the mirror symmetry implies that every CY manifold has a partner with an opposite
Euler characteristic. We will use this fact to identify a mirror CY manifold.

Recall that the CY manifold $M$ is of type $\mathbb{A}$ while the other CY manifold $\widetilde{M}$
is of type $\mathbb{B}$. Thus the spin connection of $M \; (\widetilde{M})$ acts on the spinor vector
space $S_+ = \mathbf{4} \; (S_- = \overline{\mathbf{4}})$ of positive (negative) chirality.
Since the classes $\mathbb{A}$ and $\mathbb{B}$ are completely independent and
defined in the different vector spaces, one can choose the pair $(M, \widetilde{M})$ such that
their Euler characteristics satisfy the relation $\chi(M) = - \chi(\widetilde{M})$
and so the mirror relation \eq{mirror-cyh1}. Let us explain why this is possible.

Every complex vector bundle $E$ of rank $n$ has an underlying real vector bundle
$E_\mathbb{R}$ of rank $2n$, obtained by discarding the complex structure on each fiber.
Then the top Chern class of a complex vector bundle $E$ is the Euler class of
its realization \ct{bott-tu}
\begin{equation}\label{chern-euler}
    c_n(E) = e(E_\mathbb{R})
\end{equation}
where $n = \mathrm{rank} \; E$.
Therefore, the Euler characteristic $\chi(M)$ of $M$
for a tangent bundle $E_\mathbb{R} = TM$ is given by the integral of
the top Chern class
\begin{equation}\label{euler-top}
    \chi(M) = \int_M c_n(E).
\end{equation}
Recall that if $E$ is a complex vector bundle, then there exists
a dual or conjugate bundle $\overline{E}$ with an opposite complex structure
whose $j$-th Chern class is given by \ct{spin-book,bott-tu}
\begin{equation}\label{dual-chern}
    c_j(\overline{E}) = (-1)^j c_j(E).
\end{equation}

The Euler characteristic $\chi(M)$ for a six-dimensional Riemannian manifold $M$
is given by
\begin{eqnarray}\label{euler}
\chi(M) &\equiv& - \frac{1}{ 2^7 \cdot 3 \pi^3} \int_M
\varepsilon^{A_1 A_2 \cdots A_6}
R_{A_1A_2} \wedge  R_{A_3A_4} \wedge R_{A_5A_6} \xx
&=& - \frac{1}{2^{10} \cdot 3 \pi^3}
\int_M d^6 x \varepsilon^{M_1 M_2 \cdots M_6}
\varepsilon^{A_1 A_2 \cdots A_6} R_{M_1 M_2 A_1 A_2}
R_{M_3 M_4 A_3 A_4} R_{M_5 M_6 A_5 A_6}.
\end{eqnarray}
On one hand, for the type $\mathbb{A}$ in Eq. \eq{id-fa} where $R^{(+)}_{AB} = F^{(+)a}\eta^a_{AB}$,
it is given by
\begin{eqnarray}\label{euler-a}
\chi_+(M) &=& - \frac{1}{ 2^7 \cdot 3 \pi^3} \int_M
\varepsilon^{A_1 A_2 \cdots A_6}
R^{(+)}_{A_1A_2} \wedge  R^{(+)}_{A_3A_4} \wedge R^{(+)}_{A_5A_6} \xx
&=& - \frac{1}{ 2^{10} \cdot 3 \pi^3} \int_M
(\varepsilon^{A_1 A_2 \cdots A_6}  \eta^a_{A_1A_2} \eta^b_{A_3A_4} \eta^c_{A_5A_6})
F^{(+)a} \wedge F^{(+)b}  \wedge F^{(+)c} \xx
&=& - \frac{1}{ 96 \pi^3} \int_M d^{abc}
F^{(+)a} \wedge F^{(+)b}  \wedge F^{(+)c}
\end{eqnarray}
where Eq. \eq{eta-3} was used. On the other hand, for the type $\mathbb{B}$ in Eq. \eq{id-fb}
where $R^{(-)}_{AB} = F^{(-)a}\overline{\eta}^a_{AB}$,
the Euler characteristic \eq{euler} can be written as
\begin{eqnarray}\label{euler-b}
\chi_-(\widetilde{M}) &=& \frac{1}{ 2^7 \cdot 3 \pi^3} \int_{\widetilde{M}}
\varepsilon^{A_1 A_2 \cdots A_6}
R^{(-)}_{A_1A_2} \wedge  R^{(-)}_{A_3A_4} \wedge R^{(-)}_{A_5A_6} \xx
&=& \frac{1}{ 96 \pi^3} \int_{\widetilde{M}} d^{abc}
F^{(-)a} \wedge F^{(-)b}  \wedge F^{(-)c}
\end{eqnarray}
where Eq. \eq{eta-4} was used.\footnote{\label{sign-euler}In order to define
the Euler characteristic for the type $\mathbb{B}$, it is considered that the flip of chirality
corresponds to the parity transformation (see appendix A) and so the orientation reversal.
That is the reason for the sign flip of $\chi_-(\widetilde{M})$.
But there is some ambiguity for the choice of sign because the six-dimensional
Euler characteristic need not be positive unlike the four-dimensional case. This sign ambiguity is insignificant
since it can be compensated with the redefinition $R^{(-)}_{AB} \to - R^{(-)}_{AB}$.
Hence one may keep the same sign convention for $\chi_+ (M)$ and $\chi_-(\widetilde{M})$.
In any case the mirror pair $(M, \widetilde{M})$ will be defined
by the condition $\chi_+ (M) = - \chi_-(\widetilde{M})$.
The doubling of geometric variety guarantees the freedom to arrange a mirror pair $(M, \widetilde{M})$
to satisfy the relation $\chi_+ (M) = - \chi_-(\widetilde{M})$.}
It is straightforward to represent the above Euler characteristics
in terms of the chiral bases \eq{dec-fa} and \eq{dec-fb}.
For the type $\mathbb{A}$ where $ F^{(+)a} = f^{ab}_{(++)} J^b_+$,
$\chi_+(M)$ using the identity \eq{vol-id} reads as
\begin{equation}\label{euler-su4+}
\chi_+(M) = - \frac{1}{192 \pi^3} \int_M d^6 x \sqrt{g^{(+)}}
d^{abc} d^{def} f^{ad}_{(++)}
f^{be}_{(++)} f^{cf}_{(++)}.
\end{equation}
Similarly, $\chi_-(\widetilde{M})$ for the type $\mathbb{B}$ where $F^{(-)a} = f^{ab}_{(--)} J^b_-$
can be written as
\begin{equation}\label{euler-su4-}
\chi_-(\widetilde{M}) = \frac{1}{192 \pi^3} \int_{\widetilde{M}} d^6 x \sqrt{g^{(-)}}
d^{abc} d^{def} f^{ad}_{(--)}
f^{be}_{(--)} f^{cf}_{(--)}.
\end{equation}

Recall that two irreducible spinor representations of $Spin(6)$ can be identified
with the fundamental and anti-fundamental representations of $SU(4)$.
By choosing a complex structure, the $Spin(6)$ tangent bundle $TM$ reduces to a $U(3)$ vector bundle $E$.
In order to utilize the relation \eq{euler-top}, let us consider the $U(3) \subset SU(4)$
sub-bundle $E$ such that $TM \otimes \mathbb{C} = E \oplus \overline{E}$.
Note that the $U(3)$ does not mix an underlying complex structure.
Thus we embed the class $\mathbb{A}$ into the $U(3)$ vector bundle $E$ over $M$.
Similarly, by considering the complexification $T\widetilde{M} \otimes \mathbb{C} = F \oplus \overline{F}$,
the class $\mathbb{B}$ is embedded into the $U(3)$ vector bundle $\overline{F}$ over $\widetilde{M}$.
It is important to recall that the curvature coefficients $f^{ab}_{(++)}$ and $f^{ab}_{(--)}$
are determined by completely independent metrics $g^{(+)}$ on $M$ and
$g^{(-)}$ on $\widetilde{M}$, respectively.
Therefore it is always possible to find a pair $(M, \widetilde{M})$
such that the Euler characteristics \eq{euler-su4+} for the type  $\mathbb{A}$
and \eq{euler-su4-} for the type $\mathbb{B}$ have a precisely opposite sign, i.e.,
$\chi_+(M)=-\chi_-(\widetilde{M})$. For a CY manifold $M$ whose holonomy
is $SU(3)$, the structure constants $d^{abc}$ take values only in the $su(3) \subset u(3)$ Lie algebra.
In this case, the Euler characteristic $\chi(M)$ is given by \ct{string-book}
\begin{equation}\label{euler-cy}
    \chi(M) = 2 \big(h^{1,1} (M) - h^{2,1} (M) \big).
\end{equation}
Considering the definition of the Hodge number $h^{p,q} (M) = \mathrm{dim}H^{p,q}(M) \geq 0$,
the sign flip of the Euler characteristics, $\chi_+(M) = -\chi_-(\widetilde{M})$,
can be explained if the pair $(M, \widetilde{M})$ satisfy the mirror relation
\begin{equation}\label{mirror-ab}
h^{1,1}(\mathbb{A}) = h^{2,1}(\mathbb{B}), \qquad h^{1,1}(\mathbb{B}) = h^{2,1}(\mathbb{A}).
\end{equation}
Indeed the mirror relation (\ref{mirror-ab}) is the only way to explain the sign flip of the Euler
characteristic.

The mirror symmetry \eq{mirror-ab} can be further clarified by using the fact that
the Euler characteristic $\chi(M)$ of a spin manifold $M$ is related to the index
of the Dirac operator on $M$ \ct{vacuum-string}. Denote the Dirac index for fermion fields
in a representation $R$ by $index(R)$.
The Euler characteristic $\chi(M)$ is then given by
\begin{equation}\label{mirror-index}
    \chi(M) = index(R) - index(\overline{R})
\end{equation}
where $\overline{R}$ is the complex conjugate representation of $R$.
Let $\mathbf{4}$ be the fundamental representation of $SU(4)$ and
$\overline{\mathbf{4}}$ its complex conjugate, i.e. the anti-fundamental representation.
Then $index(\overline{\mathbf{4}}) = - index(\mathbf{4})$ since in six dimensions
the complex conjugation exchanges positive and negative chirality zero modes while also
exchanging $\mathbf{4}$ and $\overline{\mathbf{4}}$. Under the $SU(3)$ representation,
$\mathbf{4} = \mathbf{1} \oplus \mathbf{3}$ and $\overline{\mathbf{4}} = \overline{\mathbf{1}}
\oplus \overline{\mathbf{3}}$ where $index(\mathbf{1}) =
index(\overline{\mathbf{1}}) = 0$, so the Euler characteristic \eq{mirror-index}
is given by \ct{vacuum-string}
\begin{eqnarray}\label{mirror-index3}
    \chi(M) &=& index(\mathbf{4}) - index(\overline{\mathbf{4}}) \xx
    &=& 2 \, index(\mathbf{4}) \xx
    &=&  index(\mathbf{3}) - index(\overline{\mathbf{3}}) \xx
    &=& 2 \, index(\mathbf{3}).
\end{eqnarray}
Then the identity \eq{mirror-index3} immediately implies the relation $\chi_+(M) = -
\chi_-(\widetilde{M})$ for a pair of spin manifolds embedded in the opposite chirality
representations $\mathbf{4}$ and  $\overline{\mathbf{4}}$ (or $\mathbf{3}$
and $\overline{\mathbf{3}}$ for CY manifolds).
This result is consistent with the mirror symmetry \eq{mirror-ab}
since $\mathbb{A} \cong \mathbf{3}$ and $\mathbb{B} \cong \overline{\mathbf{3}}$.

\section{Mirror Symmetry from Gauge Theory}

In section 3, the six-dimensional Euclidean gravity has been formulated as
$SU(4) \cong Spin(6)$ Yang-Mills gauge theory. It was shown that
a K\"ahler manifold is described by the reduced $U(3) \subset SU(4)$ gauge symmetry.
After imposing the Ricci-flat condition on the K\"ahler manifold,
the gauge group in the Yang-Mills theory is further reduced to $SU(3)$.
After all, a CY manifold $M$ from the gauge theory point of view can be described by $SU(3)$ connections
supported on $M$ satisfying the HYM equations.
And the mirror symmetry says that a CY manifold has a mirror pair satisfying the relation \eq{mirror-cyh1}.
Therefore, there must be a corresponding HYM instanton which can be derived from a mirror CY manifold
obeying the mirror relation \eq{mirror-ab}. In this section we will identify the mirror HYM instanton
from the gauge theory approach and then clarify the mirror symmetry between CY manifolds
from the gauge theory formulation.

Suppose that the metric of a six-dimensional Riemannian manifold $M$ is given by
\be \la{back-space}
ds^2 = g_{MN}(x) dx^M dx^N.
\ee
Let $\pi: E \to M$ be an $SU(4)$ bundle over $M$ whose curvature is defined by
\bea \la{field-str}
F &=& dA + A \wedge A  \xx
&=& \half \Big( \partial_M A_N - \partial_N A_M + [A_M, A_N] \Big) dx^M \wedge dx^N
\eea
where $A = A^a_M (x) T^a dx^M$ is a connection one-form on the vector bundle $E$.
The generators $T^a$ of $su(4)$ Lie algebra satisfy the commutation
relation \eq{lie-algebra} with normalization $\mathrm{Tr} T^a T^b
= - \frac{1}{2} \delta^{ab}$. Consider the $SU(4)$ Yang-Mills gauge theory defined
on the Riemannnian manifold $M$ with the metric \eq{back-space} whose action is given by
\begin{equation}\label{curved-ym}
    S_{YM} = - \frac{1}{2 g^2_{YM}} \int_{M} d^6 x \sqrt{g} g^{MP}g^{NQ}
    \mathrm{Tr} F_{MN} F_{PQ}.
\end{equation}
Using the projection operator \eq{so6-pro-op} and the identity \eq{proj-genid},
it is easy to derive the following formula
\begin{eqnarray} \la{po-identity}
\big(P_\pm F \big)^2 & = & \big(P_\pm^{A_1B_1A_2B_2} F_{A_2B_2} \big)
\big(P_\pm^{A_1B_1A_3B_3} F_{A_3B_3} \big) \xx
   &=& \frac{1}{4} \Big( F_{A_1B_1} \pm \frac{1}{4} \varepsilon^{A_1B_1A_2B_2A_3B_3}
  F_{A_2B_2} I_{A_3B_3} \Big)^2 \xx
  &=& \frac{1}{2} F_{AB} F^{AB} \pm \frac{1}{8} \varepsilon^{ABCDEF} F_{AB} F_{CD} I_{EF}
  + \frac{1}{8} \big(I_{AB} F^{AB} \big)^2 \xx
  &=& P_\pm^{ABCD} F_{AB} F_{CD} + \frac{1}{8}
  \big(I_{AB} F^{AB} \big)^2.
\end{eqnarray}
One can rewrite the action \eq{curved-ym} using the above identity as
\begin{eqnarray}\label{bps-cym}
    S_{YM} &=& - \frac{1}{4 g^2_{YM}} \int_{M} d^6 x  \sqrt{g}
    \mathrm{Tr} \left[\Big( F_{AB} \pm \frac{1}{4} \varepsilon^{ABCDEF}
  F_{CD} I_{EF} \Big)^2 - \frac{1}{2}
  \big(I_{AB} F^{AB} \big)^2 \right. \xx
  && \hspace{4cm} \left. \mp \frac{1}{2} \varepsilon^{ABCDEF} F_{AB} F_{CD} I_{EF}
  \right].
\end{eqnarray}
The above action can be written in a more compact form as
\begin{eqnarray}\label{bps-ym}
    S_{YM} &=&  - \frac{1}{4 g^2_{YM}} \int_{M} d^6 x \, \sqrt{g} \,
    \mathrm{Tr} \left[\Big( F_{AB} \pm * (F \wedge \Omega \big)_{AB} \Big)^2
    - \frac{1}{2} \big(I_{AB} F^{AB} \big)^2 \right] \xx
  && \pm \frac{1}{g^2_{YM}} \int_{M} d^6 x \mathrm{Tr}
  F \wedge F \wedge \Omega
\end{eqnarray}
where $\Omega$ is the two-form of rank 6 defined by Eq. \eq{kahler-form}.

Using the fact
\begin{equation}\label{cs-term}
    \mathrm{Tr} F \wedge F = d \, \mathrm{Tr} \big( A \wedge F -\frac{1}{3}
    A \wedge A \wedge A \big) \equiv d K,
\end{equation}
one can see that the last term in Eq. \eq{bps-ym} is a topological term, i.e.,
\begin{equation}\label{top-term}
    \mathrm{Tr} F \wedge F \wedge \Omega = d (K \wedge \Omega)
\end{equation}
if and only if the two-form $\Omega$ is closed, i.e. $d\Omega=0$. In other words,
when $M$ is a K\"ahler manifold, the last term in Eq. \eq{bps-ym} depends only
on the topological class of the K\"ahler-form $\Omega$ and the vector bundle $E$ over $M$.
Note that Eq. \eq{bps-ym}, except the second term, is very similar to the Bogomol'nyi equation
for Yang-Mills instantons whose action is bounded by a topological term.
Indeed we can apply the Bogomol'nyi argument to Eq. \eq{bps-ym} thanks to
the identity $\frac{1}{8} \varepsilon^{ABCDEF} I_{CD} I_{EF} = I_{AB}$.
More precisely, it is easy to see that a solution obeying the $\Omega$-self-duality equations
\be \la{duy-ym}
F_{AB} \pm * (F \wedge \Omega \big)_{AB} = 0
\ee
automatically satisfies the condition
\begin{equation}\label{stability}
  I_{AB} F^{AB} = 0.
\end{equation}
Therefore the minimum action can be achieved by the configuration satisfying
Eq. \eq{duy-ym} and is given by the last term--the topological term--in Eq. \eq{bps-ym}.
Note that we have already encountered the above self-duality equations in Eqs. \eq{omega-asd}
and \eq{omega+asd}. They can be summarized as the so-called DUY equations \cite{duy}
\bea \la{duy-ym1}
&& F^{(2,0)} = F^{(0,2)} = 0, \\
\la{duy-ym2}
&& F \wedge \Omega^2 = 0.
\eea
The first equation states that the $SU(4)$ gauge field is a connection on a holomorphic vector bundle
and the last condition corresponds to the stability of the holomorphic vector bundle in algebraic geometry.
It is straightforward to show \ct{string-book,gang-tian} that a solution of the self-duality equations \eq{duy-ym}
automatically satisfies the Yang-Mills equations of motion
\begin{equation}\label{ym-eq}
    g^{MN} D_M F_{NP} = 0
\end{equation}
on a K\"ahler manifold.

Let us analyze the HYM equations \eq{duy-ym}.
We observed in section 3 that the 't Hooft symbols in Eq. \eq{6-thooft} realizes the isomorphism
between irreducible $spin(6)$ Lorentz algebra and $su(4)$ Lie algebra and provides a complete basis
of two-forms in $\Omega^2_\pm (M)$.
For instance, one may expand the $SU(4)$ field strengths $F^a_{AB} \; (a = 1, \cdots, 15)$
using the basis \eq{6-thooft} like either Eq. \eq{dec-fa} or \eq{dec-fb}.
A question is how to realize the doubling of CY manifolds from the $SU(4)$ gauge theory approach.
The crux for this question is that the $N$-dimensional fundamental representation of $SU(N)$
for $N$ greater than two is a complex representation, whose complex conjugate is often called
the anti-fundamental representation.
And the complex conjugate representation $\overline{\mathbf{N}}$ is an inequivalent representation
different from the original one $\mathbf{N}$. In particular, the positive and negative chirality
representations of $Spin(6) \cong SU(4)$ coincide with the fundamental $(\mathbf{4})$
and the anti-fundamental $(\overline{\mathbf{4}})$ representations of $SU(4)$.
Therefore we have a freedom to embed the solutions of Yang-Mills gauge theory in a specific representation.
This freedom is basically related to the existence of two independent bases of two-forms,
$\eta^a_{AB}$ and $\overline{\eta}^a_{AB}$, according to the isomorphism (B) and (C).

Thereby we will identify the $SU(4)$ field strength $F^a_{AB}$ in the fundamental representation
$\mathbf{4}$ with the type $\mathbb{A}$ in Eq. \eq{dec-fa} and in the anti-fundamental representation $\overline{\mathbf{4}}$ with the type $\mathbb{B}$ in Eq. \eq{dec-fb}.
For the anti-fundamental representation $\overline{\mathbf{4}}$,
the Lie algebra generators are given by $(T^a)^* =  - \frac{i}{2} \lambda^*_a$
and they obey the same Lie algebra as $T^a$:
\begin{equation}\label{su4*}
    [(T^a)^*, (T^b)^*] = - f^{abc} (T^c)^*.
\end{equation}
But one can see from \eq{su4-product} that the symmetric structure constants have an opposite sign, i.e.,
\begin{equation}\label{dd*}
   \Tr \{T^a, T^b\} T^c = - \frac{i}{2} d^{abc}, \qquad
     \Tr \{(T^a)^*, (T^b)^*\} (T^c)^* = \frac{i}{2} d^{abc}.
\end{equation}
It turns out that this sign flip is correlated with the opposite sign in Eq. \eq{anti-comm}.
According to the tensor product \eq{tensor-prod},
one can decompose the coefficients $f^{ab}_{(\pm\pm)}$ into a symmetric part
and an antisymmetric part
\begin{equation}\label{gen-dec-f}
    f^{ab}_{(\pm\pm)} = f^{(ab)}_{(\pm\pm)} + f^{[ab]}_{(\pm\pm)}.
\end{equation}
Although it is not necessary to impose the symmetry property \eq{symm-coef}
for a general vector bundle $\pi: E \to M$, we will impose the symmetric prescription,
i.e. $f^{[ab]}_{(\pm\pm)}=0$, because we are interested in the gauge theory formulation
of six-dimensional Riemannian manifolds where $E = TM$ and the bundle connections are
identified with spin connections.\footnote{Of course,
the symmetric condition \eq{symm-dec} greatly reduces the number of field strengths
$(225 \to 120)$. According to the relation \eq{64} for the tangent bundle $E=TM$,
it is easy to derive the identities $[F^{(\pm)}_{AB}, J^{AB}_\pm]=0$ and
$\frac{1}{4} \{ F^{(\pm)}_{AB}, J^{AB}_\pm \} = - \frac{1}{8}f^{ab}_{(\pm\pm)} \delta^{ab} +
\frac{i}{2} d^{abc} f^{ab}_{(\pm\pm)} T_\pm^c$.}
Then,
\begin{equation}\label{symm-dec}
 F^{(\pm)a}_{AB} = f^{ab}_{(\pm\pm)} \eta^{(\pm)b}_{AB},
\end{equation}
where we have omitted the symmetrization symbol with respect to $a \leftrightarrow b$
for brevity.

Now let us consider the HYM equations on a K\"ahler manifold $M$.
Recall that the HYM equation \eq{duy-ym} can be resolved by
decomposing the Yang-Mills field strengths \eq{symm-dec} into
the eigenspaces of the Hodge operator \eq{w-hodge}. And we showed that
the decomposition \eq{2-from-dec} is equivalent to the branching \eq{branching1} and \eq{branching2}
of $SU(4)$ under the $U(3)$ subgroup since the $SU(4)$ gauge group is reduced to $U(3)$
by the background K\"ahler class $\Omega$.\footnote{It might be obvious from the expansion \eq{symm-dec}
which intertwines the $SU(4)$ index $a$ and $SO(6)$ indices $A,B$. Note that the different choice of
background K\"ahler classes can be parameterized by the homogeneous space $SU(4)/U(3) = \mathbb{C}P^3$.
Also Eq. \eq{complex-thooft} implies that the space $\mathbb{C}P^3 = SU(4)/U(3)$ can be identified
with the space of complex structure deformations \ct{spin-book}.
This coincidence might presage the mirror symmetry.}
Therefore, the Yang-Mills field strengths obeying Eq. \eq{duy-ym} take values in $u(3)$ Lie algebra,
i.e., $a,b$ run over $3,6,7,8,11,12,13,14,15$ for the fundamental representation $\mathbf{4}$
while $3,4,5,8,9,10,13,14,15$ for the anti-fundamental representation $\overline{\mathbf{4}}$.
To be specific, $\eta^a_{AB} \in \{l^{(+)\hat{a}}_{AB}, n^{(+)0}_{AB} \}$ for $\mathbf{4}$
and $\overline{\eta}^a_{AB} \in \{l^{(-)\hat{a}}_{AB}, n^{(-)0}_{AB} \}$ for
$\overline{\mathbf{4}}$ with the 't Hooft symbols $l^{(\pm)\hat{a}}_{AB}$ and $n^{(\pm)0}_{AB}$
defined in section 3. As the background K\"ahler class $\Omega$ determines a particular
$U(3) \subset SU(4)$ subgroup and $\mathbf{4}$ and $\overline{\mathbf{4}}$ belong to two different
representations, the K\"ahler classes in the representations $\mathbf{4}$ and $\overline{\mathbf{4}}$
should be attributed to different K\"ahler manifolds.

Hence let us consider the $SU(4)$ gauge theory defined on two different K\"ahler manifolds $M$ and $\widetilde{M}$
whose background K\"ahler classes are, respectively, given by
\begin{equation}\label{kahler-44*}
    \Omega_{+} = n^{(+)0}_{AB} e^{(+)A} \wedge e^{(+)B}, \qquad
    \Omega_{-} = n^{(-)0}_{AB} e^{(-)A} \wedge e^{(-)B}.
\end{equation}
Given a fixed K\"ahler class, the HYM equations will be solved by $U(3)$ connections.
The stability equation \eq{stability} for each case is then reduced to the following equations
\begin{eqnarray}\label{stability+eq}
    && I_{AB} F^{(+)a}_{AB} = f^{ab}_{(++)} \eta^b_{AB} I_{AB} = 0
    \quad \Leftrightarrow \quad  f_{(++)}^{3a} + \frac{1}{\sqrt{3}} f_{(++)}^{8a}
    + \frac{1}{\sqrt{6}} f_{(++)}^{15a} = 0, \\
    \label{stability-eq}
    && I_{AB} F^{(-)a}_{AB} = f^{ab}_{(--)} \overline{\eta}^b_{AB}
    I_{AB} = 0 \quad \Leftrightarrow \quad
     f_{(--)}^{3a} - \frac{1}{\sqrt{3}} f_{(--)}^{8a} - \frac{1}{\sqrt{6}} f_{(--)}^{15a} = 0.
\end{eqnarray}
By applying the exactly same argument as section 3, one can conclude that the above equations
are equivalent to the vanishing of the first Chern-class, i.e.,
\begin{equation}\label{dd11}
     F^{(+)0} = d A^{(+)0} = 0,  \qquad F^{(-)0} = d A^{(-)0} = 0,
\end{equation}
where the $U(1)$ gauge fields $A^{(\pm)0}$ are defined by Eqs. \eq{u1-gauge} and \eq{-u1-gauge}.
One can also see from Eq. \eq{omega-asd} that the $SU(3)$ basis
$\{l^{(\pm)\hat{a}}_{AB}\}$ definitely picks up the $+$-sign in Eq. \eq{duy-ym} and
its solution is given by
\begin{equation}\label{duy-++}
   F^{(\pm)\hat{a}}_{AB} = f^{\hat{a}\hat{b}}_{(\pm\pm)} l^{(\pm)\hat{b}}_{AB},
   \qquad \hat{a}, \hat{b} = 1, \cdots, 8.
\end{equation}

Consequently we found that the HYM instanton inherited from a CY manifold is described by the $SU(3)$ connections
with the trivial first Chern class. This means that the tangent bundle $TM$ of a CY manifold $M$
gives rise to $SU(3)$ connections in a {\it stable} holomorphic vector bundle \ct{string-book}.
This is exactly the statement of the DUY theorem \cite{duy} for a particular case of
the vector bundle $E=TM$ over a CY manifold $M$.
Now it becomes clear what is the mirror relation for the HYM instantons.
The mirror symmetry of CY manifolds can be understood as the relationship between
two kinds of HYM instantons in $SU(4)$ gauge theory embedded in the fundamental representation $\mathbf{4}$
and the anti-fundamental representation $\overline{\mathbf{4}}$.
Each representation has its own cohomology classes, taking values in the holomorphic vector bundlee $E$
over $M$ and $\overline{F}$ over $\widetilde{M}$.
It should be remarked that we intend to construct the complex vector
bundles $E$ and $\overline{F}$ via the tangent bundles $TM \otimes \mathbb{C} = E \oplus \overline{E}$
and $T\widetilde{M} \otimes \mathbb{C} = F \oplus \overline{F}$, so the vector
bundles $E$ and $\overline{F}$ are independent of each other.
Since the K\"ahler class $\Omega_\pm$ reduces the gauge group to $U(3)$,
the underlying complex structures are not mixed under gauge transformations.

Since we want to understand the mirror symmetry between CY manifolds in terms of
$SU(4)$ gauge theory, it will be useful to calculate the Chern classes of the vector bundle
to elucidate the mirror symmetry between HYM instantons.
It was already shown that the first Chern class $c_1(E)$ of the holomorphic vector bundle
satisfying Eq. \eq{duy-ym} is trivial, i.e. $c_1(E) = 0$.
Also we have shown that the last term in Eq. \eq{bps-ym} is a topological invariant which
contains the second Chern class $c_2(E)$.
After using Eq. \eq{duy-ym}, one can derive the inequality
\begin{equation} \la{bogo-ineq}
  \frac{1}{8 \pi^2} \int_{M_\pm} \mathrm{Tr}
  F^{(\pm)} \wedge F^{(\pm)} \wedge \Omega_{\pm} \geq 0,
\end{equation}
where
\begin{equation}\label{2-chern}
    c_2 (V_\pm) = \frac{1}{8 \pi^2} \mathrm{Tr} F^{(\pm)} \wedge F^{(\pm)}
\end{equation}
is the second Chern class of a complex vector bundle $V_+ = E$ over $M_+ = M$ or
$V_- = \overline{F}$ over $M_- = \widetilde{M}$.
This is known as the Bogomolov inequality \ct{gang-tian,bogomolov},
which is true for all stable bundles with $c_1(V_\pm) = 0$.
Using the identification in Eqs. \eq{id-fa} and \eq{id-fb},
one may translate the above inequality into the one in gravity theory
\begin{equation}\label{2-chern-gravity}
 - \frac{1}{16 \pi^2} \int_{M_\pm}  R^{(\pm)}_{AB} \wedge R^{(\pm)}_{AB} \wedge
  \Omega_\pm \geq 0.
\end{equation}

Finally, according to the formula (\ref{euler-top}), we calculate the integral of the third Chern
class $c_3 (V_\pm)$ given by
\begin{eqnarray}\label{3-chern-a}
    \chi_+(E) &=& - \frac{i}{24 \pi^3} \int_{M} \mathrm{Tr}
  F^{(+)} \wedge F^{(+)} \wedge F^{(+)} \xx
  &=& - \frac{1}{96 \pi^3} \int_{M} d^{abc}
  F^{(+)a} \wedge F^{(+)b} \wedge F^{(+)c} \xx
  &=& - \frac{1}{192 \pi^3} \int_M d^6 x \sqrt{g^{(+)}}
d^{abc} d^{def} f^{ad}_{(++)}
f^{be}_{(++)} f^{cf}_{(++)}
\end{eqnarray}
and
\begin{eqnarray}\label{3-chern-b}
    \chi_-(\overline{F}) &=& - \frac{i}{24 \pi^3} \int_{\widetilde{M}} \mathrm{Tr}
  F^{(-)} \wedge F^{(-)} \wedge F^{(-)} \xx
  &=& \frac{1}{96 \pi^3} \int_{\widetilde{M}} d^{abc}
  F^{(-)a} \wedge F^{(-)b} \wedge F^{(-)c} \xx
  &=& \frac{1}{192 \pi^3} \int_{\widetilde{M}} d^6 x \sqrt{g^{(-)}}
d^{abc} d^{def} f^{ad}_{(--)}
f^{be}_{(--)} f^{cf}_{(--)}.
\end{eqnarray}
It might be remarked that the relative sign $c_3(\overline{F}) = - c_3(F)$
for the third Chern classes of a complex vector bundle $F$ and its conjugate bundle $\overline{F}$
arises from the property \eq{dd*}.
Since the complex vector bundles $E$ and $\overline{F}$ are independently defined over
two different K\"ahler manifolds $M$ and $\widetilde{M}$, respectively,
the expansion coefficients $f^{ab}_{(\pm\pm)}$ in Eq. (\ref{symm-dec}) will also be separately
determined by them. Hence it should be possible to construct a pair of complex vector bundles
$(E \to M, \overline{F} \to \widetilde{M})$ such that $\chi_+(E) = - \chi_-(\overline{F})$.
One may notice that the sign flip in the Euler characteristic is
also consistent with the general result \eq{mirror-index3}.
In consequence, the above Euler characteristics correctly reproduce Eqs. \eq{euler-su4+} and \eq{euler-su4-}
for the CY manifold $M$ and its mirror manifold $\widetilde{M}$.
This constitutes a gauge theory formulation of mirror symmetry.

In conclusion we have confirmed the picture depicted in the Table \ref{table-ms} that
the mirror symmetry between CY manifolds can be understood as the mirror pair of
HYM instantons in the fundamental representations
$\mathbf{3}$ and $\overline{\mathbf{3}}$ of $SU(3)$ gauge connections.
Since the existence of two different fundamental representations of $SU(4) \cong Spin (6)$
is related to the doubling of the vector space in Eq. \eq{doubling-6vec}
according to the isomorphism \eq{clifford-ext}, we see that the mirror symmetry of CY manifolds and
HYM instantons originates from the Hodge duality in the vector space $\Lambda^* M$.

\section{Discussion}

The physics on a curved spacetime becomes more transparent when expressed in a locally
inertial frame and it is even indispensable when one wants to couple spinors to gravity since spinors
in $d$-dimensions form a representation of $Spin(d)$ Lorentz group rather than $GL(d, \mathbb{R})$.
It is also required to take an irreducible spinor representation of the Lorentz symmetry.
Then one can apply the elementary propositions (A,B,C) in section 1 to $d$-dimensional Riemannian
manifolds to see their consequences. It is, especially, interesting to apply them
to six-dimensional CY manifolds.
The proposition (A) first says that Riemann curvature tensors $R_{ABCD}$ carry two kinds of indices;
the first group, say $[AB]$, belongs to $Spin(6)$ indices and the second group $[CD]$ belongs to
form indices in $\Omega^2(M)$. But the proposition (C) requires that two groups must have an isomorphic
structure as vector spaces. After imposing the torsion free condition that leads to
the symmetry property, $R_{ABCD} = R_{CDAB}$, the vector space structure for the two groups
should be even identified. For example, the irreducible spinor representation of
the Lorentz group $Spin(6)$ requires us to consider the vector spaces $\Omega^2 (M)$ and $\Omega^4 (M)$
on an equal footing. The doubling of the vector space \eq{doubling-6vec}
is realized as either two independent chiral representations of the Lorentz group $Spin(6)$ or
two independent complex representations $\mathbf{4}$ and $\overline{\mathbf{4}}$ of the gauge group $SU(4)$.
We observed that the doubling of the vector spaces essentially brings about the doubling for the variety of
six-dimensional spin manifolds which is responsible for the existence of the mirror symmetry
between CY manifolds.

It may be worthwhile to compare the four and six dimensions in perspective. On a four-dimensional orientable
manifold, the vector space of two-forms $\Omega^2 (M)$ is not doubled because the Hodge-dual of a two-form
is again a two-form. Instead the vector space $\Omega^2 (M)$ splits canonically into two vector spaces
as (\ref{2-form-dec}). This split is resonant with the self-duality of chiral Lorentz
generators $J_\pm^{AB}$ because they obey the relation
\begin{equation}\label{4d-sdj}
   J_\pm^{AB} = \pm \frac{1}{2} \varepsilon^{ABCD} J_\pm^{CD}.
\end{equation}
Therefore the chiral Lorentz generators $J_\pm^{AB}$ have three independent components only.
Combining them together, they consist of six generators which match with the dimension of $\Omega^2 (M)$.
Applying this fact to Eq. \eq{riemann-tensor}, one can see that $R^{(\pm)}_{AB}$ contains $18= 6 \times 3$
components and so $36 = 18 + 18$ components in total, which is the number of components of Riemann
curvature tensors $R_{ABCD}$ before imposing the first Bianchi identity.
This situation is different from the six-dimensional case as is evident from the comparison of
Eqs. (\ref{doubling-6vec}) and  (\ref{2-form-dec}).
This difference is originated from the fact that, in six dimensions, there is another source of
two-forms coming from the Hodge-dual of four-forms.
As a consequence, $R^{(\pm)}_{AB}$ in Eq. (\ref{proj+lmn})
has $225 = 15 \times 15$ components in six dimensions and so $450 = 225 + 225$ components in total
before imposing the first Bianchi identity. After imposing the first Bianchi identity in each class
that totally comprises $240 = 120 + 120$ constraints,
the physical curvature tensors $\big( R^{(+)}_{AB} \oplus R^{(-)}_{AB} \big)$ have $210 = 105 + 105$
components in total. This doubling for the variety of CY manifolds is a core origin
of the mirror symmetry between CY manifolds.

Via the gauge theory formulation of six-dimensional Euclidean gravity,
we showed that HYM instantons can be constructed in two different ways by embedding them into
the fundamental or anti-fundamental representation of $SU(4) \cong Spin(6)$ gauge group.
Since a CY manifold can be recast as a HYM instanton from the gauge theory point of view
(see the quotation in section 1) and the chiral representation of $Spin(6)$
corresponds to the fundamental representation of $SU(4)$,
the structure in the Table \ref{table-ms} has been nicely verified.
After all, the mirror symmetry of CY manifolds can be understood as the existence of
the mirror pair of HYM instantons by doubling the variety of six-dimensional spin manifolds
according to the Hodge duality \eq{doubling-6vec}.

Strominger, Yau and Zaslow recently proposed \ct{syz-mirror} that the mirror symmetry
is a T-duality transformation along a dual special Lagrangian tori fibration
on a mirror CY manifold. The T-duality transformation along the dual three-tori
introduces a sign flip in the Euler characteristic as even and odd forms exchange their role.
Note that the odd number of T-duality operations
transforms type IIB string theory to type IIA string theory and vice versa.
Hence the type IIA and IIB CY manifolds will be mirror to each other because the six-dimensional
chirality will be flipped after the T-duality and the ten-dimensional chirality is correlated
with the six-dimensional one. This result implies that the mirror symmetry in string theory originates
from the two different chiral representations of CY manifolds, which is consistent with our picture.

Our gauge theory formulation may be generalized to a general six-dimensional
Riemannian manifold like the four-dimensional case \ct{ohya,loy} because it is
simply based on the general propositions $(A, B, C)$ in section 1.
One may consider, for example, the Strominger system \ct{strominger,li-yau}
for non-K\"ahler complex manifolds. The Strominger system admits a conformally balanced Hermitian
form on a three-dimensional compact complex manifold $M$, a nowhere vanishing
holomorphic $(3, 0)$-form and a HYM connection on a vector bundle $E$ over
this manifold. The consistency of the underlying physical theory imposes a constraint
that the curvature forms have to satisfy the anomaly equation.
As far as the non-K\"ahler CY manifold admits a spin structure,
the gauge theory formulation for the Strominger system may be straightforward
as much as we have done in this paper. Thus it may be interesting to formulate the mirror symmetry
for non-K\"ahler manifolds from the gauge theory perspective and to generalize it to the case
without spin structure, for instance, a manifold with $Spin^{\mathbb{C}}$ structure only.
If there is a substantial progress along this line, it will be reported elsewhere.

If we consider a CY manifold $M$ to be the HYM instanton of the tangent bundle $TM$,
this instanton will have their own moduli space given by their zero modes, with the non-zero modes
providing various ``uplifting". Then the following questions naturally arise:
How is the moduli space of HYM instantons related to the CY moduli space?
Also how do the non-zero modes of the instanton solution correspond to the CY deformations?

To discuss this issue, let us consider an infinitesimal deformation of the gauge field
\begin{equation}\label{def-a}
A_\mu + \delta A_\mu.
\end{equation}
If we demand Eq. \eq{duy-ym1} for both the original gauge field and the deformation,
then the deformation must satisfy $\overline{\partial} \delta A = 0$.
This means that $\delta A \in H^1 (\mathrm{End} \, E)$.
On a manifold of $SU(3)$ holonomy and for the case $E = TM$,
$H^1 (\mathrm{End} \, E)$ coincides with $H^{2,1}(M)$.
Therefore the bundle moduli (\ref{def-a}) for the condition \eq{duy-ym1} of
the holomorphic tangent bundle $TM$ correspond to the deformations of the complex structure,
counted by $H^1(M, TM)\cong H^{2,1}(M)$ \cite{vacuum-string,mirror-book}.
However it is known \cite{s-wall} that the HYM equations \eq{duy-ym1} and \eq{duy-ym2} do not fix any of
the K\"ahler moduli. Indeed the DUY theorem states \cite{duy} that, {\it for a fixed choice of K\"ahler moduli},
there exists a solution of the HYM equations if the holomorphic vector bundle is slope-stable,
which is the case for the tangent bundle. Therefore it is not possible to extract the K\"ahler moduli
from the bundle moduli (\ref{def-a}). This implies that the moduli space of a CY manifold is not
fully captured by the instanton moduli space even for the tangent bundle
since we fix the K\"ahler moduli of a background CY manifold to define the HYM equations.

It is well-known that the instanton moduli space has singularities, the so-called ``small instanton" singularities.
Also the CY moduli space has singularities: the conifold points \cite{mirror-book}.
Thus it will be interesting to understand how these two kinds of singularities are related to each other.
Our result implies that the singularities in the instanton moduli space
are related to the singularities (the conifold points) of the CY moduli space because
the tangent bundle $TM$ is defined by the CY manifold $M$.
But the blown-up of the conifold singularities may arise in different ways from the instanton picture
of the CY manifold since it is known \cite{conifold} that there is a natural complex structure on the resolution
but not a natural K\"ahler structure while the deformation is symplectic in a natural way but not naturally complex.
Since only the complex structure deformations of the CY manifold are encoded in the bundle moduli \eq{def-a},
we speculate that the resolved conifold is realized from the instanton side while the deformed conifold appears
in the background CY manifold. Note that the deformed conifold is mirror to the resolved conifold,
which is related by the conifold transitions. Thus the instanton picture of CY manifolds implies that
the HYM instanton for the tangent bundle $TM$ over a deformed conifold $M$ is mirror to the HYM instanton
for the tangent bundle $T\widetilde{M}$ over a resolved conifold $\widetilde{M}$. We leave this problem
for the future work.

Mirror symmetry provides an isomorphism between complex geometry and symplectic geometry which
relates the deformation of complex structure on the complex geometry side to the counting
of pseudo-holomorphic spheres on the symplectic geometry side. A very similar picture arises
in emergent gravity that isomorphically relates the deformation of symplectic structure
described by a NC $U(1)$ gauge theory to the deformation of complex structure in Einstein gravity.
The deformation of symplectic structure is represented by $\mathcal{F} = B + F$ where $B$ is
an underlying symplectic structure on $M$ and $F = dA$ is identified with the curvature of
line bundle $L \to M$. In order to allow singular $U(1)$ gauge fields such as $U(1)$ instantons,
it is necessary to generalize the line bundle to a torsion free sheaf or an ideal sheaf.
NC $U(1)$ gauge fields are introduced via a local coordinate transformation
$\phi \in \mathrm{Diff}(M)$ eliminating $U(1)$ gauge fields, i.e., $\phi^* (\mathcal{F}) = B$,
known as the Darboux theorem in symplectic geometry. It was claimed in \cite{hsy11,hsy13} and
recently shown in \cite{hsy14} that six-dimensional CY manifolds are emergent from NC Hermitian
$U(1)$ instantons. Note that the NC Hermitian $U(1)$ instantons correspond to $U(1)$ connections
in a {\it stable} holomorphic line bundle $L \to M$ or more generally a torsion free sheaf
(an ideal sheaf) from the commutative description \cite{nlinst}.
When we conceive the emergent CY manifolds from the mirror symmetry perspective,
an interesting question is how to realize the mirror symmetry from
the emergent gravity picture. It turns out \cite{mirror-emg} that the emergent gravity picture
provides a very nice result for the mirror symmetry.

\section*{Acknowledgments}

We are grateful to Chanju Kim for a collaboration at an early stage and
very helpful discussions. HSY thanks Richard Szabo for his warm hospitality and valuable discussions
during his visit to Heriot-Watt University, Edinburgh.
He also thanks Jungjai Lee and John J. Oh for constructive discussions and collaborations
with a related subject.
The work of H.S. Yang was supported by the National Research Foundation of Korea (NRF) grant funded
by the Korea government (MOE) (No. NRF-2015R1D1A1A01059710).
The work of S. Yun was supported by the World Class University grant number R32-10130.

\newpage

\appendix

\section{$Spin(6)$ and $SU(4)$}

We consider the six-dimensional Clifford algebra with the Dirac matrices given by
\begin{equation}\label{6d-dirac}
    \Gamma^A = \left(
                 \begin{array}{cc}
                   0 & \gamma^A \\
                   \overline{\gamma}^A & 0 \\
                 \end{array}
               \right), \qquad A=1, \cdots, 6
\end{equation}
where $\overline{\gamma}^A = (\gamma^A)^\dagger$. Thus the Dirac matrices we have taken are Hermitian, i.e.,
$(\Gamma^A)^\dagger = \Gamma^A$. We choose $(\gamma^i)^\dagger = \gamma^i \; (i=1, \cdots, 5)$
and $(\gamma^6)^\dagger = - \gamma^6$.
We will use the following representation of Dirac matrices \cite{aru-fro}
\begin{eqnarray} \la{dirac-matrix}
&& \gamma^1 = \left(
  \begin{array}{cccc}
    0 & 0 & 0 & -1 \\
    0 & 0 & 1 & 0 \\
    0 & 1 & 0 & 0 \\
    -1 & 0 & 0 & 0 \\
  \end{array}
\right), \quad
\gamma^2 = \left(
  \begin{array}{cccc}
    0 & 0 & 0 & i \\
    0 & 0 & i & 0 \\
    0 & -i & 0 & 0 \\
    -i & 0 & 0 & 0 \\
  \end{array}
\right), \quad
\gamma^3 = \left(
  \begin{array}{cccc}
    0 & 0 & 1 & 0 \\
    0 & 0 & 0 & 1 \\
    1 & 0 & 0 & 0 \\
    0 & 1 & 0 & 0 \\
  \end{array}
\right), \xx
&& \gamma^4 = \left(
  \begin{array}{cccc}
    0 & 0 & -i & 0 \\
    0 & 0 & 0 & i \\
    i & 0 & 0 & 0 \\
    0 & -i & 0 & 0 \\
  \end{array}
\right), \quad
\gamma^5 = \left(
  \begin{array}{cccc}
    1 & 0 & 0 & 0 \\
    0 & 1 & 0 & 0 \\
    0 & 0 & -1 & 0 \\
    0 & 0 & 0 & -1 \\
  \end{array}
\right) = - \gamma^1 \gamma^2 \gamma^3 \gamma^4,
\end{eqnarray}
satisfying the $Spin(5)$ Clifford algebra relation
\begin{equation}\label{so5-clifford}
    \gamma^i \gamma^j + \gamma^j \gamma^i = 2 \delta^{ij}, \qquad i,j=1, \cdots, 5.
\end{equation}
The Lorentz generators for the irreducible (chiral) spinor representation of $Spin(6)$
are defined by
\begin{equation}\label{so6+gen}
    J_\pm^{AB} \equiv \frac{1}{2}(\mathbf{I}_8 \pm \Gamma_7) J^{AB}
\end{equation}
where $\Gamma_7 = i \Gamma^1 \cdots \Gamma^6$. Note that $J_+^{AB}$ and $J_-^{AB}$
independently satisfy the Lorentz algebra \eq{lorentz-algebra} and commute each other, i.e.,
$[J_+^{AB}, J_-^{CD}] = 0$. They also satisfy the anti-commutation relation
\begin{equation}\label{anti-comm}
    \{J_\pm^{AB}, J_\pm^{CD} \} = -\frac{1}{2} \big( \delta^{AC} \delta^{BD} -
 \delta^{AD} \delta^{BC} \big) \Gamma_\pm \pm \frac{i}{2} \varepsilon^{ABCDEF} J_\pm^{EF}.
\end{equation}
Because the chiral matrix $\Gamma_7$ is given by
\begin{equation}\label{gamma7}
    \Gamma_7 = \left(
                 \begin{array}{cc}
                   \mathbf{I}_4 & 0 \\
                   0 & - \mathbf{I}_4 \\
                 \end{array}
               \right)
\end{equation}
where $\mathbf{I}_4$ is the $4 \times 4$ identity matrix, the generators of the chiral spinor representation
in Eq. \eq{so6+gen} are given by $4 \times 4$ matrices.
Then the two independent chiral spinor representations of $Spin(6)$ are given by
\begin{eqnarray} \la{spin+rep}
  J_+^{AB} &=& \{ J_+^{ij} = \frac{1}{4}[\gamma^i, \gamma^j], \; J_+^{i6}
  = \frac{i}{2} \gamma^i \}, \\
  \la{spin-rep}
  J_-^{AB} &=& \{ J_-^{ij} = \frac{1}{4}[\gamma^i, \gamma^j], \; J_-^{i6}
  = - \frac{i}{2} \gamma^i \}.
\end{eqnarray}
One can check that the generators $J_+^{AB}$ and $J_-^{AB}$ separately obey the
Lorentz algebra \eq{lorentz-algebra}.

One can exchange the positive chiral representation and the negative chiral representation
by a parity transformation, which is a reflection $x^M \to - x^M$ of any one element of
the fundamental six-dimensional representation of $Spin(6)$ \ct{string-book}; in our case, $x^6 \to - x^6$.
But they cannot be connected by any $SO(6)$ rotations.

The anti-Hermitian $4 \times 4$ matrices $T^a = \frac{i}{2} \lambda_a, \; a = 1, \cdots, 15$
with vanishing traces constitute the basis of $SU(4)$ Lie algebra. The Hermitian $4 \times 4$
matrices $\lambda_a$ are given by
\begin{eqnarray} \la{su4-matrix}
&& \lambda_1 = \left(
              \begin{array}{cccc}
                0 & 1 & 0 & 0 \\
                1 & 0 & 0 & 0 \\
                0 & 0 & 0 & 0 \\
                0 & 0 & 0 & 0 \\
              \end{array}
            \right), \quad
            \lambda_2 = \left(
              \begin{array}{cccc}
                0 & -i & 0 & 0 \\
                i & 0 & 0 & 0 \\
                0 & 0 & 0 & 0 \\
                0 & 0 & 0 & 0 \\
              \end{array}
            \right), \quad
            \lambda_3 = \left(
              \begin{array}{cccc}
                1 & 0 & 0 & 0 \\
                0 & -1 & 0 & 0 \\
                0 & 0 & 0 & 0 \\
                0 & 0 & 0 & 0 \\
              \end{array}
            \right), \xx
  &&  \lambda_4 = \left(
              \begin{array}{cccc}
                0 & 0 & 1 & 0 \\
                0 & 0 & 0 & 0 \\
                1 & 0 & 0 & 0 \\
                0 & 0 & 0 & 0 \\
              \end{array}
            \right), \quad
            \lambda_5 = \left(
              \begin{array}{cccc}
                0 & 0 & -i & 0 \\
                0 & 0 & 0 & 0 \\
                i & 0 & 0 & 0 \\
                0 & 0 & 0 & 0 \\
              \end{array}
            \right), \quad
            \lambda_6 = \left(
              \begin{array}{cccc}
                0 & 0 & 0 & 0 \\
                0 & 0 & 1 & 0 \\
                0 & 1 & 0 & 0 \\
                0 & 0 & 0 & 0 \\
              \end{array}
            \right), \xx
  &&  \lambda_7 = \left(
              \begin{array}{cccc}
                0 & 0 & 0 & 0 \\
                0 & 0 & -i & 0 \\
                0 & i & 0 & 0 \\
                0 & 0 & 0 & 0 \\
              \end{array}
            \right), \quad
            \lambda_8 = \frac{1}{\sqrt{3}}\left(
              \begin{array}{cccc}
                1 & 0 & 0 & 0 \\
                0 & 1 & 0 & 0 \\
                0 & 0 & -2 & 0 \\
                0 & 0 & 0 & 0 \\
              \end{array}
            \right), \quad
            \lambda_9 = \left(
              \begin{array}{cccc}
                0 & 0 & 0 & 1 \\
                0 & 0 & 0 & 0 \\
                0 & 0 & 0 & 0 \\
                1 & 0 & 0 & 0 \\
              \end{array}
            \right), \xx
             &&  \lambda_{10} = \left(
              \begin{array}{cccc}
                0 & 0 & 0 & -i \\
                0 & 0 & 0 & 0 \\
                0 & 0 & 0 & 0 \\
                i & 0 & 0 & 0 \\
              \end{array}
            \right), \quad
            \lambda_{11} = \left(
              \begin{array}{cccc}
                0 & 0 & 0 & 0 \\
                0 & 0 & 0 & 1 \\
                0 & 0 & 0 & 0 \\
                0 & 1 & 0 & 0 \\
              \end{array}
            \right), \quad
            \lambda_{12} = \left(
              \begin{array}{cccc}
                0 & 0 & 0 & 0 \\
                0 & 0 & 0 & -i \\
                0 & 0 & 0 & 0 \\
                0 & i & 0 & 0 \\
              \end{array}
            \right), \xx
             &&  \lambda_{13} = \left(
              \begin{array}{cccc}
                0 & 0 & 0 & 0 \\
                0 & 0 & 0 & 0 \\
                0 & 0 & 0 & 1 \\
                0 & 0 & 1 & 0 \\
              \end{array}
            \right), \quad
            \lambda_{14} = \left(
              \begin{array}{cccc}
                0 & 0 & 0 & 0 \\
                0 & 0 & 0 & 0 \\
                0 & 0 & 0 & -i \\
                0 & 0 & i & 0 \\
              \end{array}
            \right), \quad
            \lambda_{15} = \frac{1}{\sqrt{6}}\left(
              \begin{array}{cccc}
                1 & 0 & 0 & 0 \\
                0 & 1 & 0 & 0 \\
                0 & 0 & 1 & 0 \\
                0 & 0 & 0 & -3 \\
              \end{array}
            \right).
\end{eqnarray}
The generators satisfy the following relation
\begin{equation} \label{su4-product}
    T^a T^b = - \frac{1}{8} \delta^{ab} \mathbf{I}_4 - \frac{1}{2} f^{abc} T^c
    + \frac{i}{2} d^{abc} T^c
\end{equation}
where the structure constants $f^{abc}$ are completely antisymmetric while $d^{abc}$
are symmetric with respect to all of their indices. Their values are shown up in the Tables 2 and 3.
We have got these tables from Ref. \ct{greiner}.

\begin{center}
\begin{table}[ht]
\centering
\begin{tabular}{c c c c | c c c c | c c c c}
  \hline
  $a$ & $b$ & $c$ & $f^{abc}$ & $a$ & $b$ & $c$ & $f^{abc}$ & $a$ & $b$ & $c$ & $f^{abc}$ \\
  \hline
  1 & 2 & 3 & 1 & 3 & 6 & 7 & $-\frac{1}{2}$ & 6 & 12 & 13 & $-\frac{1}{2}$ \\
  1 & 4 & 7 & $\frac{1}{2}$ & 3 & 9 & 10 & $\frac{1}{2}$ & 7 & 11 & 13 & $\frac{1}{2}$ \\
  1 & 5 & 6 & $-\frac{1}{2}$ & 3 & 11 & 12 & $-\frac{1}{2}$ & 7 & 12 & 14 & $\frac{1}{2}$ \\
  1 & 9 & 12 & $\frac{1}{2}$ & 4 & 5 & 8 & $\frac{\sqrt{3}}{2}$ & 8 & 9 & 10 & $\frac{1}{2\sqrt{3}}$ \\
  1 & 10 & 11 & $-\frac{1}{2}$ & 4 & 9 & 14 & $\frac{1}{2}$ & 8 & 11 & 12 & $\frac{1}{2\sqrt{3}}$ \\
  2 & 4 & 6 & $\frac{1}{2}$ & 4 & 10 & 13 & $-\frac{1}{2}$ & 8 & 13 & 14 & $-\frac{1}{\sqrt{3}}$ \\
  2 & 5 & 7 & $\frac{1}{2}$ & 5 & 9 & 13 & $\frac{1}{2}$ & 9 & 10 & 15 & $\sqrt{\frac{2}{3}}$ \\
  2 & 9 & 11 & $\frac{1}{2}$ & 5 & 10 & 14 & $\frac{1}{2}$ & 11 & 12 & 15 & $\sqrt{\frac{2}{3}}$ \\
  2 & 10 & 12 & $\frac{1}{2}$ & 6 & 7 & 8 & $\frac{\sqrt{3}}{2}$ & 13 & 14 & 15 & $\sqrt{\frac{2}{3}}$ \\
  3 & 4 & 5 & $\frac{1}{2}$ & 6 & 11 & 14 & $\frac{1}{2}$ & \\
  \hline
\end{tabular}
\caption{The nonvanishing structure constants $f^{abc}$}
\label{table-sf}
\end{table}
\end{center}

\begin{center}
\begin{table}[ht]
\centering
\begin{tabular}{c c c c | c c c c | c c c c}
  \hline
  $a$ & $b$ & $c$ & $d^{abc}$ & $a$ & $b$ & $c$ & $d^{abc}$ & $a$ & $b$ & $c$ & $d^{abc}$ \\
  \hline
  1 & 1 & 8 & $\frac{1}{\sqrt{3}}$ & 3 & 9 & 9 & $\frac{1}{2}$ & 7 & 11 & 14 & $ -\frac{1}{2}$   \\
  1 & 1 & 15 & $\frac{1}{\sqrt{6}}$ & 3 & 10 & 10 & $\frac{1}{2}$ & 7 & 12 & 13 & $\frac{1}{2}$  \\
  1 & 4 & 6 & $\frac{1}{2}$ &  3 & 11 & 11 & $-\frac{1}{2}$ & 8 & 8 & 8 & $- \frac{1}{\sqrt{3}}$  \\
  1 & 5 & 7 & $\frac{1}{2}$ & 3 & 12 & 12 & $-\frac{1}{2}$  &  8 & 8 & 15 & $\frac{1}{\sqrt{6}}$ \\
  1 & 9 & 11 & $\frac{1}{2}$ & 4 & 4 & 8 & $-\frac{1}{2\sqrt{3}}$ &  8 & 9 & 9 & $\frac{1}{2\sqrt{3}}$ \\
  1 & 10 & 12 & $\frac{1}{2}$ & 4 & 4 & 15 & $\frac{1}{\sqrt{6}}$ & 8 & 10 & 10 & $\frac{1}{2\sqrt{3}}$ \\
  2 & 2 & 8 & $\frac{1}{\sqrt{3}}$ &  4 & 9 & 13 & $\frac{1}{2}$ & 8 & 11 & 11 & $\frac{1}{2\sqrt{3}}$ \\
  2 & 2 & 15 & $\frac{1}{\sqrt{6}}$ & 4 & 10 & 14 & $\frac{1}{2}$  & 8 & 12 & 12 & $\frac{1}{2\sqrt{3}}$ \\
  2 & 4 & 7 & $-\frac{1}{2}$ & 5 & 5 & 8 & $-\frac{1}{2\sqrt{3}}$ & 8 & 13 & 13 & $-\frac{1}{\sqrt{3}}$ \\
  2 & 5 & 6 & $\frac{1}{2}$ & 5 & 5 & 15 & $\frac{1}{\sqrt{6}}$ & 8 & 14 & 14 & $-\frac{1}{\sqrt{3}}$ \\
  2 & 9 & 12 & $-\frac{1}{2}$ & 5 & 9 & 14 & $-\frac{1}{2}$ &  9 & 9 & 15 & $-\frac{1}{\sqrt{6}}$  \\
  2 & 10 & 11 & $\frac{1}{2}$ & 5 & 10 & 13 & $\frac{1}{2}$  & 10 & 10 & 15 & $- \frac{1}{\sqrt{6}}$ \\
  3 & 3 & 8 & $\frac{1}{\sqrt{3}}$ & 6 & 6 & 8 & $-\frac{1}{2\sqrt{3}}$  & 11 & 11 & 15 &
  $- \frac{1}{\sqrt{6}}$ \\
  3 & 3 & 15 & $\frac{1}{\sqrt{6}}$ & 6 & 6 & 15 & $\frac{1}{\sqrt{6}}$ &  12 & 12 & 15 &
  $-\frac{1}{\sqrt{6}}$ \\
  3 & 4 & 4 & $\frac{1}{2}$ & 6 & 11 & 13 & $\frac{1}{2}$ &  13 & 13 & 15 & $-\frac{1}{\sqrt{6}}$  \\
  3 & 5 & 5 & $\frac{1}{2}$ &  6 & 12 & 14 & $\frac{1}{2}$ & 14 & 14 & 15 & $-\frac{1}{\sqrt{6}}$ \\
  3 & 6 & 6 & $-\frac{1}{2}$ & 7 & 7 & 8 & $-\frac{1}{2\sqrt{3}}$ & 15 & 15 & 15 & $-\sqrt{\frac{2}{3}}$ \\
  3 & 7 & 7 & $-\frac{1}{2}$ & 7 & 7 & 15 & $\frac{1}{\sqrt{6}}$  \\
  \hline
\end{tabular}
\caption{The nonvanishing structure constants $d^{abc}$}
\label{table-sd}
\end{table}
\end{center}

\section{Six-dimensional 't Hooft symbols}

The explicit representation of the six-dimensional 't Hooft symbol $\eta^{a}_{AB}
= - \Tr( T^a J^{AB}_+)$ is given by
\begin{equation*}
\begin{array}{l}
\la{matrix-6thooft}
\eta^1_{AB} = \frac{1}{2} \left(
  \begin{array}{cccccc}
    0 & 0 & 0 & 1 & 0 & 0 \\
    0 & 0 & 1 & 0 & 0 & 0 \\
    0 & -1 & 0 & 0 & 0 & 0 \\
    -1 & 0 & 0 & 0 & 0 & 0 \\
    0 & 0 & 0 & 0 & 0 & 0 \\
    0 & 0 & 0 & 0 & 0 & 0 \\
  \end{array}
\right) = \frac{i}{2} \lambda_2 \otimes \sigma^1,
\end{array}
\end{equation*}

\begin{equation*}
\begin{array}{l}
\eta^2_{AB} = - \frac{1}{2} \left(
  \begin{array}{cccccc}
    0 & 0 & 1 & 0 & 0 & 0 \\
    0 & 0 & 0 & -1 & 0 & 0 \\
    -1 & 0 & 0 & 0 & 0 & 0 \\
    0 & 1 & 0 & 0 & 0 & 0 \\
    0 & 0 & 0 & 0 & 0 & 0 \\
    0 & 0 & 0 & 0 & 0 & 0 \\
  \end{array}
\right) = - \frac{i}{2} \lambda_2 \otimes \sigma^3,
\end{array}
\end{equation*}

\begin{equation*}
\begin{array}{l}
 \eta^3_{AB} = \frac{1}{2} \left(
  \begin{array}{cccccc}
    0 & 1 & 0 & 0 & 0 & 0 \\
    -1 & 0 & 0 & 0 & 0 & 0 \\
    0 & 0 & 0 & 1 & 0 & 0 \\
    0 & 0 & -1 & 0 & 0 & 0 \\
    0 & 0 & 0 & 0 & 0 & 0 \\
    0 & 0 & 0 & 0 & 0 & 0 \\
  \end{array}
\right) = \frac{i}{2} \Big(\frac{2}{3}\mathbf{I}_3
+ \frac{1}{\sqrt{3}} \lambda_8 \Big) \otimes \sigma^2,
\end{array}
\end{equation*}

\begin{equation*}
\begin{array}{l}
 \eta^4_{AB} = \frac{1}{2} \left(
  \begin{array}{cccccc}
    0 & 0 & 0 & 0 & 0 & 0\\
    0 & 0 & 0 & 0 & 0 & 0 \\
    0 & 0 & 0 & 0 & 0 & 1 \\
    0 & 0 & 0 & 0 & 1 & 0 \\
    0 & 0 & 0 & -1 & 0 & 0 \\
    0 & 0 & -1 & 0 & 0 & 0 \\
  \end{array}
\right) = \frac{i}{2} \lambda_7 \otimes \sigma^1, \\
\eta^5_{AB} = -\frac{1}{2} \left(
  \begin{array}{cccccc}
    0 & 0 & 0 & 0 & 0 & 0\\
    0 & 0 & 0 & 0 & 0 & 0 \\
    0 & 0 & 0 & 0 & 1 & 0 \\
    0 & 0 & 0 & 0 & 0 & -1 \\
    0 & 0 & -1 & 0 & 0 & 0 \\
    0 & 0 & 0 & 1 & 0 & 0 \\
  \end{array}
\right) = -\frac{i}{2} \lambda_7 \otimes \sigma^3,
\end{array}
\end{equation*}

\begin{equation*}
\begin{array}{l}
 \eta^6_{AB} = \frac{1}{2} \left(
  \begin{array}{cccccc}
    0 & 0 & 0 & 0 & 0 & 1\\
    0 & 0 & 0 & 0 & -1 & 0 \\
    0 & 0 & 0 & 0 & 0 & 0 \\
    0 & 0 & 0 & 0 & 0 & 0 \\
    0 & 1 & 0 & 0 & 0 & 0 \\
    -1 & 0 & 0 & 0 & 0 & 0 \\
  \end{array}
\right) = \frac{i}{2} \lambda_4 \otimes \sigma^2,
\end{array}
\end{equation*}

\begin{equation*}
\begin{array}{l}
\eta^7_{AB} = -\frac{1}{2} \left(
  \begin{array}{cccccc}
    0 & 0 & 0 & 0 & 1 & 0\\
    0 & 0 & 0 & 0 & 0 & 1 \\
    0 & 0 & 0 & 0 & 0 & 0 \\
    0 & 0 & 0 & 0 & 0 & 0 \\
    -1 & 0 & 0 & 0 & 0 & 0 \\
    0 & -1 & 0 & 0 & 0 & 0 \\
  \end{array}
\right) = -\frac{i}{2} \lambda_5 \otimes \mathbf{I}_2,
\end{array}
\end{equation*}

\begin{equation*}
\begin{array}{l}
\eta^8_{AB} = -\frac{1}{2\sqrt{3}} \left(
  \begin{array}{cccccc}
    0 & 1 & 0 & 0 & 0 & 0\\
    -1 & 0 & 0 & 0 & 0 & 0 \\
    0 & 0 & 0 & -1 & 0 & 0 \\
    0 & 0 & 1 & 0 & 0 & 0 \\
    0 & 0 & 0 & 0 & 0 & -2 \\
    0 & 0 & 0 & 0 & 2 & 0 \\
  \end{array}
\right) = - \frac{i}{2\sqrt{3}} \Big( -\frac{2}{3}\mathbf{I}_3
+ \lambda_3 + \frac{2}{\sqrt{3}}\lambda_8 \Big) \otimes \sigma^2,
\end{array}
\end{equation*}

\begin{equation*}
\begin{array}{l}
\eta^9_{AB} = -\frac{1}{2} \left(
  \begin{array}{cccccc}
    0 & 0 & 0 & 0 & 0 & 1\\
    0 & 0 & 0 & 0 & 1 & 0 \\
    0 & 0 & 0 & 0 & 0 & 0 \\
    0 & 0 & 0 & 0 & 0 & 0 \\
    0 & -1 & 0 & 0 & 0 & 0 \\
    -1 & 0 & 0 & 0 & 0 & 0 \\
  \end{array}
\right) = -\frac{i}{2} \lambda_5 \otimes \sigma^1,
\end{array}
\end{equation*}

\begin{equation*}
\begin{array}{l}
\eta^{10}_{AB} = \frac{1}{2} \left(
  \begin{array}{cccccc}
    0 & 0 & 0 & 0 & 1 & 0\\
    0 & 0 & 0 & 0 & 0 & -1 \\
    0 & 0 & 0 & 0 & 0 & 0 \\
    0 & 0 & 0 & 0 & 0 & 0 \\
    -1 & 0 & 0 & 0 & 0 & 0 \\
    0 & 1 & 0 & 0 & 0 & 0 \\
  \end{array}
\right) = \frac{i}{2} \lambda_5 \otimes \sigma^3,
\end{array}
\end{equation*}

\begin{equation*}
\begin{array}{l}
\eta^{11}_{AB} = \frac{1}{2} \left(
  \begin{array}{cccccc}
    0 & 0 & 0 & 0 & 0 & 0\\
    0 & 0 & 0 & 0 & 0 & 0 \\
    0 & 0 & 0 & 0 & 0 & 1 \\
    0 & 0 & 0 & 0 & -1 & 0 \\
    0 & 0 & 0 & 1 & 0 & 0 \\
    0 & 0 & -1 & 0 & 0 & 0 \\
  \end{array}
\right) = \frac{i}{2} \lambda_6 \otimes \sigma^2,
\end{array}
\end{equation*}

\begin{equation*}
\begin{array}{l}
\eta^{12}_{AB} = -\frac{1}{2} \left(
  \begin{array}{cccccc}
    0 & 0 & 0 & 0 & 0 & 0\\
    0 & 0 & 0 & 0 & 0 & 0 \\
    0 & 0 & 0 & 0 & 1 & 0 \\
    0 & 0 & 0 & 0 & 0 & 1 \\
    0 & 0 & -1 & 0 & 0 & 0 \\
    0 & 0 & 0 & -1 & 0 & 0 \\
  \end{array}
\right) = -\frac{i}{2} \lambda_7 \otimes \mathbf{I}_2,
\end{array}
\end{equation*}

\begin{equation*}
\begin{array}{l}
\eta^{13}_{AB} = \frac{1}{2} \left(
  \begin{array}{cccccc}
    0 & 0 & 0 & 1 & 0 & 0\\
    0 & 0 & -1 & 0 & 0 & 0 \\
    0 & 1 & 0 & 0 & 0 & 0 \\
    -1 & 0 & 0 & 0 & 0 & 0 \\
    0 & 0 & 0 & 0 & 0 & 0 \\
    0 & 0 & 0 & 0 & 0 & 0 \\
  \end{array}
\right) = \frac{i}{2} \lambda_1 \otimes \sigma^2,
\end{array}
\end{equation*}

\begin{equation*}
\begin{array}{l}
\eta^{14}_{AB} = \frac{1}{2} \left(
  \begin{array}{cccccc}
    0 & 0 & 1 & 0 & 0 & 0\\
    0 & 0 & 0 & 1 & 0 & 0 \\
    -1 & 0 & 0 & 0 & 0 & 0 \\
    0 & -1 & 0 & 0 & 0 & 0 \\
    0 & 0 & 0 & 0 & 0 & 0 \\
    0 & 0 & 0 & 0 & 0 & 0 \\
  \end{array}
\right) = \frac{i}{2} \lambda_2 \otimes \mathbf{I}_2,
\end{array}
\end{equation*}

\begin{equation*}
\begin{array}{l}
\eta^{15}_{AB} = \frac{1}{\sqrt{6}} \left(
  \begin{array}{cccccc}
    0 & 1 & 0 & 0 & 0 & 0\\
    -1 & 0 & 0 & 0 & 0 & 0 \\
    0 & 0 & 0 & -1 & 0 & 0 \\
    0 & 0 & 1 & 0 & 0 & 0 \\
    0 & 0 & 0 & 0 & 0 & 1 \\
    0 & 0 & 0 & 0 & -1 & 0 \\
  \end{array}
\right) = \frac{i}{\sqrt{6}} \Big( \frac{1}{3}\mathbf{I}_3
+ \lambda_3 - \frac{1}{\sqrt{3}}\lambda_8 \Big) \otimes \sigma^2,
\end{array}
\end{equation*}
where $\mathbf{I}_n$ is the $n$-dimensional identity matrix,
and $(\sigma^1,\sigma^2, \sigma^3)$ are the Pauli matrices and
$\lambda_{\hat{a}} \; (\hat{a} = 1, \cdots, 8)$ are the $SU(3)$ Gell-Mann matrices.

Another six-dimensional 't Hooft symbol $\overline{\eta}^{a}_{AB} = - \Tr \big( (T^a)^* J^{AB}_- \big)$
can be obtained similarly:
\begin{eqnarray} \la{6-thooft-b}
&& \overline{\eta}^{1}_{AB} = - \frac{i}{2} \lambda_2 \otimes \sigma^1, \quad
\overline{\eta}^{2}_{AB}= -\frac{i}{2} \lambda_2 \otimes \sigma^3, \quad
\overline{\eta}^{3}_{AB} = -\frac{i}{2} \Big(\frac{2}{3}\mathbf{I}_3
+ \frac{1}{\sqrt{3}} \lambda_8 \Big) \otimes \sigma^2, \xx
&& \overline{\eta}^{4}_{AB} = \frac{i}{2} \lambda_6 \otimes \sigma^2, \quad
\overline{\eta}^{5}_{AB} = -\frac{i}{2} \lambda_7 \otimes \mathbf{I}_2, \quad
\overline{\eta}^{6}_{AB} = \frac{i}{2} \lambda_5 \otimes \sigma^1, \xx
&& \overline{\eta}^{7}_{AB} = -\frac{i}{2} \lambda_5 \otimes \sigma^3, \quad
\overline{\eta}^{8}_{AB} = \frac{i}{2\sqrt{3}} \Big( \frac{2}{3}\mathbf{I}_3
+ \lambda_3 - \frac{2}{\sqrt{3}}\lambda_8 \Big) \otimes \sigma^2, \quad
\overline{\eta}^{9}_{AB} = - \frac{i}{2} \lambda_4 \otimes \sigma^2, \xx
&& \overline{\eta}^{10}_{AB} = \frac{i}{2} \lambda_5 \otimes \mathbf{I}_2, \quad
\overline{\eta}^{11}_{AB} = \frac{i}{2} \lambda_7 \otimes \sigma^1, \quad
\overline{\eta}^{12}_{AB} = - \frac{i}{2} \lambda_7 \otimes \sigma^3, \\
&& \overline{\eta}^{13}_{AB} = - \frac{i}{2} \lambda_1 \otimes \sigma^2, \quad
\overline{\eta}^{14}_{AB} = \frac{i}{2} \lambda_2 \otimes \mathbf{I}_2, \quad
\overline{\eta}^{15}_{AB} = \frac{i}{\sqrt{6}} \Big( \frac{1}{3}\mathbf{I}_3
- \lambda_3 - \frac{1}{\sqrt{3}}\lambda_8 \Big) \otimes \sigma^2. \nonumber
\end{eqnarray}

In order to derive the algebras obeyed by the 't Hooft symbols,
first note that either $Spin(6)$ generators $J_\pm^{AB}$
or $SU(4)$ generators $T^a_+ \equiv T^a$ and $T^a_- \equiv (T^a)^*$ can serve as a complete basis
of any traceless, Hermitian $4 \times 4$ matrix $K$, i.e.,
\begin{equation}\label{4x4basis}
K = \sum_{a=1}^{15} k^\pm_a T^a_\pm = \frac{1}{2} \sum_{A,B=1}^6 K^\pm_{AB} J_\pm^{AB}.
\end{equation}
Using the definition \eq{6-thooft}, one can easily deduce that
\begin{eqnarray} \la{46}
&& T^a_\pm = \frac{1}{2} \eta^{(\pm)a}_{AB} J^{AB}_\pm, \\
\la{64}
&& J^{AB}_\pm = 2  \eta^{(\pm)a}_{AB} T^a_\pm,
\end{eqnarray}
where $\eta^{(+)a}_{AB} \equiv \eta^{a}_{AB}$ and $\eta^{(-)a}_{AB} \equiv \overline{\eta}^{a}_{AB}$.
Then one can consider the following matrix products
\begin{eqnarray} \la{matrix-prod-1}
\mathrm{I}: && T^a_\pm T^b_\pm = \frac{1}{4} \eta^{(\pm)a}_{AB} \eta^{(\pm)b}_{CD} J^{AB}_{\pm} J^{CD}_{\pm},  \\
\la{matrix-prod-2}
\mathrm{II}: && J^{AB}_{\pm} J^{CD}_{\pm}
= 4 \eta^{(\pm)a}_{AB} \eta^{(\pm)b}_{CD} T^a_\pm T^b_\pm.
\end{eqnarray}
By applying Eqs. \eq{lorentz-algebra}, \eq{anti-comm} and \eq{su4-product} to the above matrix products,
one can easily get the algebras obeyed by the six-dimensional 't Hooft symbols:
\begin{eqnarray}
&& \eta^a_{AB} \eta^b_{AB} = \delta^{ab} = \overline{\eta}^a_{AB} \overline{\eta}^b_{AB},
\la{eta-1} \\
&& \eta^a_{AB} \eta^a_{CD} = \frac{1}{2} \big(\delta_{AC} \delta_{BD} - \delta_{AD} \delta_{BC} \big)
= \overline{\eta}^a_{AB} \overline{\eta}^a_{CD}, \la{eta-2}\\
&& \frac{1}{4} \varepsilon^{ABCDEF} \eta^a_{CD} \eta^b_{EF} = d^{abc} \eta^c_{AB}, \la{eta-3}\\
&& \frac{1}{4} \varepsilon^{ABCDEF} \overline{\eta}^a_{CD} \overline{\eta}^b_{EF} = d^{abc}
\overline{\eta}^c_{AB}, \la{eta-4}\\
&& \eta^a_{AC} \eta^b_{BC} - \eta^a_{BC} \eta^b_{AC} = f^{abc} \eta^c_{AB}, \la{eta-5}\\
&& \overline{\eta}^a_{AC} \overline{\eta}^b_{BC} - \overline{\eta}^a_{BC} \overline{\eta}^b_{AC}
= f^{abc} \overline{\eta}^c_{AB}, \la{eta-6}\\
&&  f^{abc} \eta^a_{AB} \eta^b_{CD} = \frac{1}{2} \big(\delta_{AC} \eta^c_{BD}
- \delta_{AD} \eta^c_{BC} - \delta_{BC} \eta^c_{AD} + \delta_{BD} \eta^c_{AC} \big), \la{eta-7}\\
&&  f^{abc} \overline{\eta}^a_{AB} \overline{\eta}^b_{CD}
= \frac{1}{2} \big(\delta_{AC} \overline{\eta}^c_{BD}
- \delta_{AD} \overline{\eta}^c_{BC} - \delta_{BC} \overline{\eta}^c_{AD}
+ \delta_{BD} \overline{\eta}^c_{AC} \big), \la{eta-8}\\
&&  d^{abc} \eta^a_{AB} \eta^b_{CD} = \frac{1}{4} \varepsilon^{ABCDEF} \eta^c_{EF}, \la{eta-9}\\
&&  d^{abc} \overline{\eta}^a_{AB} \overline{\eta}^b_{CD}
= \frac{1}{4} \varepsilon^{ABCDEF} \overline{\eta}^c_{EF}. \la{eta-10}
\end{eqnarray}

Finally we list the nonzero components of the 't Hooft symbols in the basis
of complex coordinates $z^\alpha = \{ z^1 = x^1 + i x^2, z^2 = x^3 + i x^4,
z^3 = x^5 + i x^6 \}$ and their complex conjugates $\bar{z}^{\bar{\alpha}}$ where
$\alpha, \bar{\alpha} = 1,2,3$. We will denote $\eta^a_{\alpha\beta}
= \eta^a_{z^\alpha z^\beta}, \; \eta^a_{\alpha\bar{\beta}}
= \eta^a_{z^\alpha \bar{z}^{\bar{\beta}}}$, etc. in the hope of no confusion
with the real basis:
\begin{eqnarray} \la{complex-thooft}
\begin{array}{llllll}
  \eta^1_{12} = -\frac{i}{4},  & \eta^2_{12} = -\frac{1}{4}, & \eta^4_{23}
  = -\frac{i}{4}, & \eta^5_{23} = -\frac{1}{4}, & \eta^9_{13}
  = \frac{i}{4}, & \eta^{10}_{13} = \frac{1}{4}, \\
  \eta^3_{1\bar{1}} = \frac{i}{4}, & \eta^3_{2\bar{2}} = \frac{i}{4}, & \eta^6_{1\bar{3}}
  = \frac{i}{4}, & \eta^7_{1\bar{3}} = - \frac{1}{4}, \\
  \eta^8_{1\bar{1}} = - \frac{i}{4\sqrt{3}}, & \eta^8_{2\bar{2}}
  = \frac{i}{4\sqrt{3}}, & \eta^8_{3\bar{3}} = \frac{i}{2\sqrt{3}}, & \eta^{11}_{2\bar{3}}
  = \frac{i}{4}, &
   \eta^{12}_{2\bar{3}} = - \frac{1}{4}, \\
 \eta^{13}_{1\bar{2}} = \frac{i}{4}, & \eta^{14}_{1\bar{2}} = \frac{1}{4},
 &  \eta^{15}_{1\bar{1}} = \frac{i}{2\sqrt{6}},  &
   \eta^{15}_{2\bar{2}} = - \frac{i}{2\sqrt{6}}, & \eta^{15}_{3\bar{3}}
   = \frac{i}{2\sqrt{6}}.
\end{array}
\end{eqnarray}
Here the complex conjugates are not shown up since they can easily be implemented.
The corresponding values of $\overline{\eta}^a_{AB}$ can be obtained from those
of $\eta^a_{AB}$ by flipping the sign for the entries $a=1,3,4,6,8,9,11,13,15$ as well as
interchanging $z^3 \leftrightarrow \bar{z}^3$ for all entries.
Note that the first line in \eq{complex-thooft} belongs to $m_{AB}^{(+)\dot{a}}$ in Eq. \eq{+257}
with purely holomorphic or anti-holomorphic indices.
This result implies that the space of complex structure deformations can
be identified with the coset space $SU(4)/U(3) = \mathbb{C}P^3$ \ct{spin-book}.


\nc{\PR}[3]{Phys. Rev. {\bf #1}, #2 (#3)}
\nc{\NPB}[3]{Nucl. Phys. {\bf B#1}, #2 (#3)}
\nc{\PLB}[3]{Phys. Lett. {\bf B#1}, #2 (#3)}
\nc{\PRD}[3]{Phys. Rev. {\bf D#1}, #2 (#3)}
\nc{\PRL}[3]{Phys. Rev. Lett. {\bf #1}, #2 (#3)}
\nc{\PREP}[3]{Phys. Rep. {\bf #1}, #2 (#3)}
\nc{\EPJ}[3]{Eur. Phys. J. {\bf C#1}, #2 (#3)}
\nc{\PTP}[3]{Prog. Theor. Phys. {\bf #1}, #2 (#3)}
\nc{\CMP}[3]{Commun. Math. Phys. {\bf #1}, #2 (#3)}
\nc{\MPLA}[3]{Mod. Phys. Lett. {\bf A#1}, #2 (#3)}
\nc{\CQG}[3]{Class. Quant. Grav. {\bf #1}, #2 (#3)}
\nc{\NCB}[3]{Nuovo Cimento {\bf B#1}, #2 (#3)}
\nc{\ANNP}[3]{Ann. Phys. (N.Y.) {\bf #1}, #2 (#3)}
\nc{\GRG}[3]{Gen. Rel. Grav. {\bf #1}, #2 (#3)}
\nc{\MNRAS}[3]{Mon. Not. Roy. Astron. Soc. {\bf #1}, #2 (#3)}
\nc{\JHEP}[3]{J. High Energy Phys. {\bf #1}, #2 (#3)}
\nc{\JCAP}[3]{JCAP {\bf #1}, #2 {#3}}
\nc{\ATMP}[3]{Adv. Theor. Math. Phys. {\bf #1}, #2 (#3)}
\nc{\AJP}[3]{Am. J. Phys. {\bf #1}, #2 (#3)}
\nc{\ibid}[3]{{\it ibid.} {\bf #1}, #2 (#3)}
\nc{\ZP}[3]{Z. Physik {\bf #1}, #2 (#3)}
\nc{\PRSL}[3]{Proc. Roy. Soc. Lond. {\bf A#1}, #2 (#3)}
\nc{\LMP}[3]{Lett. Math. Phys. {\bf #1}, #2 (#3)}
\nc{\AM}[3]{Ann. Math. {\bf #1}, #2 (#3)}
\nc{\hepth}[1]{{\tt [arXiv:hep-th/{#1}]}}
\nc{\grqc}[1]{{\tt [arXiv:gr-qc/{#1}]}}
\nc{\astro}[1]{{\tt [arXiv:astro-ph/{#1}]}}
\nc{\hepph}[1]{{\tt [arXiv:hep-ph/{#1}]}}
\nc{\phys}[1]{{\tt [arXiv:physics/{#1}]}}
\nc{\arxiv}[1]{{\tt [arXiv:{#1}]}}


\end{document}